\documentclass[lettersize,journal]{IEEEtran}
\usepackage{amsmath,amsfonts}
\usepackage{algorithmic}
\usepackage{array}
\usepackage[caption=false,font=normalsize,labelfont=sf,textfont=sf]{subfig}
\usepackage{textcomp}
\usepackage{stfloats}
\usepackage{url}
\usepackage{verbatim}
\usepackage{graphicx}
\usepackage{cite}

\usepackage[normalem]{ulem}

\usepackage{multirow}
\usepackage{booktabs}
\usepackage{xcolor}
\usepackage{color}

\usepackage{tikz}
\usepackage{pgfplots}

\usepackage{lipsum}
\usepackage{graphicx}
\usepackage{subfig}

\usepackage[utf8]{inputenc} 
\usepackage[T1]{fontenc}

\usepackage[ruled, vlined, linesnumbered]{algorithm2e}

\usepackage{amsthm}
\newtheorem{thm}{Theorem}

\newtheorem{remark}{Remark}

\definecolor{fig_6_color_1}{RGB}{200,36,35}
\definecolor{fig_6_color_2}{RGB}{248,172,140}
\definecolor{fig_6_color_3}{RGB}{154,201,219}
\definecolor{fig_6_color_4}{RGB}{40,120,181}

\definecolor{fig_7_color_1}{RGB}{199,109,162}
\definecolor{fig_7_color_2}{RGB}{137,131,191}
\definecolor{fig_7_color_3}{RGB}{5,185,226}
\definecolor{fig_7_color_4}{RGB}{50,184,151}

\definecolor{fig_8_color_1}{RGB}{73,108,136}
\definecolor{fig_8_color_2}{RGB}{254,178,180}

\begin{document}

\title{A Unified Framework for Exploratory Learning-Aided Community Detection Under Topological Uncertainty}

\author{Yu Hou,~\IEEEmembership{}
        Cong Tran,~\IEEEmembership{}
        Ming Li,~\IEEEmembership{Member,~IEEE,}
        and Won-Yong Shin,~\IEEEmembership{Senior Member,~IEEE}
\IEEEcompsocitemizethanks{\IEEEcompsocthanksitem Y. Hou is with the School of Mathematics and Computing (Computational Science and Engineering), Yonsei University, Seoul 03722, Republic of Korea (E-mail: houyu@yonsei.ac.kr).
\IEEEcompsocthanksitem C. Tran is with the Faculty of Information Technology, Posts and Telecommunications Institute of Technology, Hanoi 100000, Vietnam (E-mail: congtt@ptit.edu.vn).
\IEEEcompsocthanksitem M. Li is with Zhejiang Institute of Optoelectronics, Jinhua 321004, China, and also with Zhejiang Key Laboratory of Intelligent Education Technology and Application, Zhejiang Normal University, Jinhua 321004, China (E-mail: mingli@zjnu.edu.cn).
\IEEEcompsocthanksitem W.-Y. Shin is with the School of Mathematics and Computing (Computational Science and Engineering), Yonsei University, Seoul 03722, Republic of Korea, and the Graduate School of Artificial Intelligence, Pohang University of Science and Technology (POSTECH), Pohang 37673, Republic of Korea (E-mail: wy.shin@yonsei.ac.kr).\\
{\it (Corresponding author: Won-Yong Shin.)}
}

}



\maketitle

\begin{abstract}
In social networks, the discovery of community structures has received considerable attention as a fundamental problem in various network analysis tasks. However, due to privacy concerns or access restrictions, the network structure is often {\em uncertain}, thereby rendering established community detection approaches ineffective without costly network topology acquisition. To tackle this challenge, we present \textsf{META-CODE}, a unified framework for detecting overlapping communities via {\em exploratory learning} aided by {\em easy-to-collect} node metadata when networks are topologically unknown (or only partially known). Specifically, \textsf{META-CODE} consists of three iterative steps in addition to the initial network inference step: 1) node-level {\em community-affiliation embeddings} based on graph neural networks (GNNs) trained by our new reconstruction loss, 2) {\em network exploration} via community-affiliation-based node queries, and 3) {\em network inference} using an edge connectivity-based Siamese neural network model from the explored network. Through extensive experiments on three real-world datasets including two large networks, we demonstrate: (a) the superiority of \textsf{META-CODE} over benchmark community detection methods, achieving remarkable gains up to 65.55\% on the Facebook dataset over the best competitor among our selected competitive methods in terms of normalized mutual information (NMI), (b) the impact of each module in \textsf{META-CODE}, (c) the effectiveness of node queries in \textsf{META-CODE} based on empirical evaluations and theoretical findings, and (d) the convergence of the inferred network.
\end{abstract}


\section{Introduction}
\label{sec:introduction}

\subsection{Background and Motivation}

\IEEEPARstart{C}{ommunity} detection \cite{su2022comprehensive, abbe2017community} is one of critical and important tasks in various network analyses to understand the fundamental features of networks. In social networks, community memberships are often allowed to overlap, wherein nodes belong to multiple communities at once \cite{palla2005uncovering}. Consequently, community detection algorithms need to be designed to accommodate this inherently overlapping nature, and as such, a range of techniques such as clique- and modularity-based approaches \cite{palla2005uncovering, nicosia2009extending}, non-negative matrix factorization (NMF)-based approaches \cite{yang2013overlapping, ye2018deep}, and deep learning-based approaches \cite{shchur2019overlapping, mehta2019stochastic} have been developed to detect overlapping communities.

While the aforementioned community detection methods always rely on the network structure, the underlying {\it true} network is {\it initially unknown or only partially known} in various domains of real-world network applications \cite{valente2007identifying}. Thus, downstream machine learning (ML) tasks on graphs, including community detection, are ineffective unless the network structure is available. Although one could invest additional efforts into uncovering the complete network prior to downstream applications, collecting complete topological information is prohibitively expensive and labor-intensive \cite{valente2007identifying}. Alternatively, one could leverage network data collected from social media to {\it infer} the unknown relationships among nodes in the underlying network. Despite the increasing availability of social media platforms, this source of information is not without limitations, such as restricted resources and more stringent privacy settings specified by users. For example, a demographic analysis of Facebook users in New York City conducted in June 2011 revealed that 52.6\% of users hid their friend list \cite{dey2012facebook}. Moreover, network connectivity information often includes many weak (uncertain) links \cite{granovetter1973strength}, such as outdated friendships, which are not effective at detecting communities. Consequently, as long as network analyses are concerned, one should assume that none or only a part of the network structure is available in practice \cite{kim2011network, tran2020sf, tran2021community}. In real-world scenarios, such incomplete network information hinders various practical applications that rely on community detection. For instance, health agencies aiming to design targeted HIV-intervention strategies often have only partial knowledge of how individuals are connected due to privacy restrictions, which hinders community detection by limiting the ability to accurately identify key subgroups for targeted interventions \cite{rice2012mobilizing}. Similarly, corporate teams seeking to improve collaborative efficiency may have limited data regarding employees' social ties, making it difficult to identify underlying subgroups essential for knowledge sharing \cite{cross2002making}. Hence, community detection must contend with a lack of full topological information.

On one hand, there have been several attempts to solve ML problems in networks with unknown topology. For instance, {\em exploratory} influence maximization \cite{wilder2018end, kamarthi2019influence, tran2024meta} was introduced along with a solution to information retrieval over unknown networks by querying individual nodes to explore their neighborhoods. Despite the success of exploratory influence maximization in solving the problem of influence maximization, the research related to {\it community detection} via exploratory learning remains {\it unexplored} yet. On the other hand, in the setting where the network structure is {\it partially observable}, there have been attempts to solve the problem of community detection by reconstructing the unobserved part of the underlying network from the observed part and then detecting the community structure \cite{yan2011finding, tran2021community}. Although there has been a considerable amount of research on community detection, most existing methods more or less require the network structure to infer the community structure, which can be labor-intensive in practical situations. 

In our study, we focus primarily on the case where the potentially existing network structure is {\it entirely unknown}, treating the scenario of partially observable networks as a special case, for the following reasons. First, performing overlapping community detection with a fully unknown (uncertain) topology is significantly more challenging due to the lack of ML tools that operate without structural information. It is worth noting that, as summarized in Table \ref{tab:literature}, {\bf the setting of networks with fully unknown topology poses distinct technical difficulties that must be approached differently from the implementation of community detection methods designed for partially observable networks.} Second, unlike the prior studies, our topologically unknown setting, naturally accommodates the problem of detecting communities when the underlying network is partially observable; in other words, solving our problem innately leads to a solution to the community detection in networks with partially known topology as well (which will be specified later).

\begin{table*}[t]
\centering
\captionsetup{skip=0pt}
\caption{Summary of the difference among overlapping community detection methods with respect to the degree of topological information and the existence of exploratory learning.}
\label{tab:literature}
\centering
\footnotesize
\begin{tabular}{l c c c c c c c}
\toprule
\multirow{2}{*}{Methods} & \multicolumn{3}{c}{Required information} & \multirow{2}{*}{Exploratory learning}  \\ \cmidrule(l){2-4} 
& Full topology  & Partial topology & Metadata \\ 
\cmidrule(r){1-5}

AGM \cite{yang2012community}, BIGCLAM \cite{yang2013overlapping}, DANMF \cite{ye2018deep}, vGraph \cite{sun2019vgraph} & \checkmark  &     & \\  
CESNA \cite{yang2013community}, NOCD \cite{shchur2019overlapping}, DGLFRM \cite{mehta2019stochastic}, PMP \cite{he2024polarized}, CAs \cite{he2021learning}, VVAMo \cite{he2021vicinal}  &  \checkmark &  & \checkmark    \\
HGP-EPM \cite{zhou2015infinite}, KroMFac \cite{tran2021community} &        &  \checkmark  &    \\ 
\textsf{META-CODE} (proposed)      &        &          &  \checkmark & \checkmark\\ 
\bottomrule
\end{tabular}
\vspace{-2.0em}
\end{table*}

In social networks, {\it node metadata} (e.g., a user's hobby or education) can be used to solve many ML problems on graphs and boost performance. Such node metadata can often be retrieved from various sources such as user-generated content \cite{leroy2010cold} or inferred from users' behavior \cite{kosinski2013private}. For example, in co-authorship networks, one can infer a topical relationship between authors by simply collecting the keywords in each author's publications. In the case of HIV awareness among homeless youth \cite{rice2012mobilizing}, metadata such as name, age, and familial background of individuals as well as homeless individuals can be easily collected through the mandatory registration process at welfare centers, required to pay monthly allowances. Because the node metadata are readily available \cite{leroy2010cold, kosinski2013private, rice2012mobilizing}, incorporating them into exploratory learning frameworks can help overcome the practical challenges associated with networks with initially unknown (or only partially known) topology. In this context, a natural question arising is: ``How to effectively perform overlapping community detection under topological uncertainty with the aid of node metadata?'' 

\subsection{Main Contributions}

To answer this question, we make the first attempt to develop an {\it overlapping} community detection method in the absence of structural information by retrieving information from both {\it node queries} and {\it node metadata}. To this end, we first outline two design challenges that must be addressed when building a new methodology.
\begin{itemize}
    \item {\bf Exploratory learning:} how to effectively {\it explore} subnetworks with the aid of {\it node queries};
    \item {\bf Exploitation of node metadata:} how to design our community detection method by maximally {\it exploiting} node metadata.
\end{itemize}

{\bf (\underline{Idea 1})} It is a quite challenging task to accurately discover community structures based solely on node metadata when network topology is initially {\it unknown}. To aid community detection, we are capable of harnessing node queries with a given query budget to retrieve structural information. Then, both the node metadata and the explored edges from node queries can be used to infer the connections between two nodes in the unexplored part, which helps identify community affiliations more precisely. In our study, we formulate a new problem that jointly finds the community-affiliation embedding matrix and a sequence of node queries in the sense that the discovered community affiliations make the inferred network and node metadata more probable.

{\bf (\underline{Idea 2})} It is worthwhile to note that existing community detection methods are no longer valid in solving the aforementioned problem (refer to Table \ref{tab:literature}). In other words, we need to present our own overlapping community detection method by establishing entirely new design principles. To achieve this goal, we propose a scalable yet effective community detection method, \textsf{META-CODE}, via {\it exploratory learning} aided by node metadata. \textsf{META-CODE} initially infers the network structure and then {\em iteratively} performs the following three steps to ensure gradually improved performance on community detection: 1) node-level {\it community-affiliation embeddings} based on graph neural networks (GNNs), widely used as a powerful means of learning graph representations, trained by our new reconstruction loss, 2) {\it network
exploration} via community-affiliation-based node queries, and 3) {\it network inference} using an edge connectivity-based Siamese neural network (EC-SiamNet) model from the explored network (refer to Fig. \ref{fig:overview} for the three steps in \textsf{META-CODE}). Our design iteratively leverages the community-affiliation embeddings to guide the query node selection strategy, ensuring that, after network exploration, the queried nodes and explored edges can enhance the inference of potential edges.

Our main contributions are summarized as follows:

\begin{itemize}
    \item {\bf Novel methodology:} We introduce \textsf{META-CODE}, a scalable end-to-end framework for solving the community detection problem in topologically uncertain networks, integrating three iterative steps—community-affiliation embedding, network exploration, and network inference—to enhance detection accuracy.

    \item {\bf Comprehensive analysis and evaluation:} We validate the rationality and effectiveness of \textsf{META-CODE} through extensive experiments on four real-world datasets including three large networks (e.g., up to x90.0 scale of the datasets used in \cite{mcauley2014discovering} in terms of the average number of nodes). We demonstrate 1) the superiority of \textsf{META-CODE} over benchmark overlapping community detection methods while showing gains of 65.55\% and 19.84\% in terms of NMI compared to the best competitor, NOCD \cite{shchur2019overlapping}, among our selected benchmark methods and a simple variant of \textsf{META-CODE}, respectively, 2) the influence of each component in \textsf{META-CODE}, and 3) the theoretical validation of our node query strategy in \textsf{META-CODE}.

\end{itemize}

\begin{table}[t]
\centering
\footnotesize
  \caption{Summary of notations.}
  \begin{tabular}{cl}
    \toprule
    \bf Notation& \bf Description\\
    \hline
    $\mathcal{G}$ & Underlying network \\
    $\mathcal{V}$ & Set of nodes in $\mathcal{G}$ \\
    $\mathcal{E}$ & Set of edges in $\mathcal{G}$ \\
    $N$ & Number of nodes \\
    $\mathcal{X}$ & Node metadata \\
    $D$ & Dimensionality of each feature vector in $\mathcal{X}$ \\
    $K$ & Number of communities \\
    $\bf F$ & Non-negative weight affiliation matrix\\
    $S_t$ & Sequence of $t$ node queries \\
    $T$ & Budget of node queries \\
    $\mathcal{E}_t$& Set of explored edges after the $t$-th query step \\
    $\mathcal{\mathcal{G}}^{\left( t \right)}$ & Inferred network after the $t$-th node query \\
    $\mathcal{E}^{\left( t \right)} $& Set of edges in $\mathcal{G}^{\left( t \right)}$ \\
    $\mathcal{E}_I^{\left( t \right)} $ & Set of inferred edges from $S_t$ and $\mathcal{E}_t $ \\
    $f_I$ & The function infers the network structure \\
    $v_t$ & The $t$-th queried node \\
    $\mathcal{N}_G(v_t)$ & Neighbors of node $v_t$ \\
    \bottomrule
\end{tabular}
\label{tab:notations}
\end{table}

\subsection{Organization}

The remainder of this paper is organized as follows. Section II is the prior studies related to our work. In Section III, we explain the methodology of our study, including the problem definition and an overview of our \textsf{META-CODE} method. Section IV describes the technical details of the proposed method. Comprehensive experiment results are shown in Section V. Finally, we provide a summary and concluding remarks in Section VI.

Table \ref{tab:notations} summarizes the notation that is used in this paper. This notation will be formally defined in the following sections when we introduce our methodology and the technical details.
 

\section{Related work}
\label{sec:RelatedWork}
Our proposed method is related to four broader areas of research, namely overlapping community detection, community detection using GNNs, community detection in incomplete networks, and network exploration.

{\bf Overlapping community detection.} Various techniques have been developed to detect overlapping communities. Approaches based on fundamental concepts such as clique and modularity were initially proposed for identifying overlapping communities \cite{palla2005uncovering, nicosia2009extending, wen2016maximal}. The community-affiliation graph model (AGM) \cite{yang2012community} was introduced as a means to detect overlapping communities by capitalizing on the observation that nodes within areas of overlapping communities tend to exhibit denser connections compared to those in non-overlapping regions. Building upon the AGM, BIGCLAM \cite{yang2013overlapping} and EPM \cite{zhou2015infinite} employed NMF and Bernoulli-Poisson  (BP) models, respectively, to detect overlapping communities. DANMF \cite{ye2018deep}, which leverages stacked NMF, and vGraph \cite{sun2019vgraph}, which utilizes a probabilistic generative model, were also presented. Additionally, a handful of methods were proposed in \cite{leskovec2012learning, yang2013community} to detect overlapping communities by incorporating the network structure and node metadata.

{\bf Community detection using GNNs.} GNNs have been widely employed to solve community detection problems \cite{zhang2019attributed, wang2021unsupervised, park2020unsupervised, fu2020magnn, choong2018learning, he2021community}. For instance, methods such as AGC \cite{zhang2019attributed} and SGCN \cite{wang2021unsupervised} leveraged graph convolution networks (GCNs) \cite{kipf2016semi} to capture complex features for community detection while aggregating neighbors' information of each node through graph convolution layers. DMGI \cite{park2020unsupervised} and MAGNN \cite{fu2020magnn} employed graph attention networks (GATs) \cite{velivckovic2017graph} to learn the importance of each node in the neighborhood adaptively. Additionally, MGAECD \cite{choong2018learning} and GUCD \cite{he2021community} utilized GCNs in autoencoder (AE)-based techniques to represent communities via embeddings in the hidden layer by reconstructing the original information. Furthermore, in \cite{sun2021graph}, a GNN encoding technique for the multi-objective evolutionary algorithm was proposed to solve the community detection problem in complex attributed networks. While the aforementioned studies have focused on detecting disjoint (i.e., non-overlapping) communities, only a few have addressed the detection of overlapping communities via GNNs \cite{shchur2019overlapping, mehta2019stochastic}. For example, NOCD \cite{shchur2019overlapping} proposed a GNN-based approach for detecting overlapping communities by fusing the network structure and node metadata, while DGLFRM \cite{mehta2019stochastic} incorporated a stochastic block model \cite{abbe2017community} with a GNN architecture for overlapping community detection. 

{\bf Community detection in incomplete networks.} Several studies have explored the problem of identifying communities when the underlying network is {\it incomplete (uncertain)} with missing edges. Some of these studies leveraged additional information, such as node metadata \cite{yang2013community} or topological similarity \cite{yan2012detecting}, for community detection. There has been a research endeavor that predicted missing edges in an incomplete network and subsequently discovered communities within the predicted topology \cite{yan2011finding}. A hierarchical gamma process infinite edge partition model \cite{zhou2015infinite} was designed not only to identify overlapping communities but also to predict missing edges. Moreover, other studies directly extracted community information from raw data sources, such as influence cascades \cite{barbieri2016efficient} and time series data \cite{hoffmann2020community}, in networks with missing edges. In \cite{tran2021community}, a particular study combined node ranking with NMF to enhance the accuracy of community detection in networks with missing edges. Table \ref{tab:literature} summarizes the difference among overlapping community detection methods when different degrees of topological information, including full topology, partial topology, and (almost) no topology, are used.

{\bf Network exploration.} The network exploration aims to explore networks whose topology is either completely unknown or partially observable to help solve downstream ML tasks on graphs. The Jump-Crawl model \cite{brautbar2010local} and BEER \cite{higashikawa2014online} were proposed as approaches for identifying influential nodes, which are used to explore the unknown network. When dealing with a network lacking topological information, numerous influence maximization methods attempted to find influential nodes and queried nodes that can retrieve their neighbors in unknown networks. Following the concept of active learning \cite{settles2009active, joshi2012scalable}, HEALER \cite{yadav2016using} investigated dynamic influence maximization over a series of rounds in which edge information is collected after each round. Additionally, exploratory influence maximization methods such as CHANGE \cite{wilder2018end} and Geometric-DQN \cite{kamarthi2019influence} were introduced by learning edges via individual node queries. IM-META \cite{tran2024meta} introduced a topology-aware ranking strategy to improve the effectiveness of network exploration in performing the influence maximization task.

\begin{remark}
\label{remark:remark1}
Although there has been a considerable amount of research on community detection \cite{yang2012community, yang2013overlapping, sun2019vgraph, zhou2015infinite, tran2021community}, most existing methods more or less require the network structure to infer the community structure, which can be labor-intensive in practical situations. Additionally, the idea of leveraging node queries and node metadata has not been explored for community detection in topologically unknown networks.
\end{remark}

\section{Methodology}
In this section, as a basis for the proposed \textsf{META-CODE} method, we first present our network model, outlining basic assumptions. We then formulate the problem of community detection in networks with unknown topology. We also provide an overview of the \textsf{META-CODE} method.

\subsection{Settings and Basic Assumptions}

Let us denote an underlying true network as $\mathcal{G} = \left( {\mathcal{V},\mathcal{E}} \right)$, where $\mathcal{V}$ is the set of $N$ nodes and $\mathcal{E}$ is the set of edges which is initially unavailable but may potentially exist. The network $\mathcal{G}$ is assumed to be an undirected unweighted {\em attributed} network without self-edges and repeated edges, having collectible node metadata ${\bf \mathcal{X}} \in \mathbb{R}^{N \times D}$ of binary-valued node features,\footnote{Non binary-valued node features can be transferred to binary values via one-hot encoding \cite{leskovec2012learning}.} where $D$ is the dimension of each feature vector. We assume that social networks follow the AGM \cite{yang2012community}, which claims that the more communities to which a pair of nodes belongs, the higher the probability that the node pair is connected. The community information can be represented by a non-negative weight affiliation matrix ${\bf F}^{(t)}\in\mathbb{R}^{N\times K}$ regarded as node-level {\it community-affiliation embeddings},\footnote{To simplify notations, ${\bf F}^{\left( t \right)} $ representing the node-level community-affiliation embeddings in the $t$-th query step will be written as ${\bf F}$ if dropping $t$ does not cause any confusion.} where $K$ is the number of communities and ${\bf F}_u$ is the affiliations of node $u$. For each node $u$, the $k$-th element ${\bf F}_{uk}$ of ${\bf F}_u$ indicates the probability that node $u$ belongs to the $k$-th community. If a certain node $u$ has multiple elements in ${\bf F}_u$ higher than a threshold,  then it belongs to multiple communities.

Similarly as in the concept of active learning \cite{settles2009active, joshi2012scalable}, we select the sequence of node queries, denoted as $S_t$, with a query budget $T$ to retrieve additional structural information from the underlying network $\mathcal{G}$, where $0 \le t \le T-1$. Note that strategies utilizing a budget of queries are widely applied to solve ML tasks on topologically unknown networks \cite{wilder2018end, kamarthi2019influence, tran2024meta}. Upon querying a single node $v_t$, we can discover its neighbors, denoted as $\mathcal{N}_\mathcal{G} \left ( v_t \right ) $, and expand the observable subnetwork accordingly, which essentially follows the same setting as prior work \cite{wilder2018end, wilder2018maximizing, tran2024meta}. Let us denote the set of explored edges after $t$ queries as $\mathcal{E}_{t}$. Then, during the $(t+1)$-th node query, we choose a node $v_{t}$ from $\mathcal{V}$ to expand and update the set of explored edges ${\cal E}_{t + 1}  = {\cal E}_t  \cup {\cal E}\left( {{\cal N}_{\cal G} \left( {v_{t} } \right),v_{t} } \right)$, where  $\mathcal{E} \left ( \mathcal{N}_\mathcal{G} \left ( v_{t} \right ),v_{t} \right )$ is a set of all edges to which each node in $\mathcal{N}_\mathcal{G} \left ( v_{t} \right ) $ and node $v_{t}$ are incident. Furthermore, the inferred network {$\mathcal{G}^{\left( t+1 \right)}$ after the $(t+1)$-th node query is represented as $ \mathcal{G}^{\left( t+1 \right)}  = \left( {\mathcal{V} ,\mathcal{E}^{\left( t+1 \right)} } \right)$, where $\mathcal{E}^{\left( t+1 \right)}$ is the set of edges in $\mathcal{G}^{\left( t+1 \right)}$ and can be updated with $\mathcal{E}^{\left( t+1 \right)}  = \mathcal{E}_{t+1}  \cup \mathcal{E}_I^{\left( t+1 \right)} $. Here, $\mathcal{E}_I^{\left( t+1 \right)}  = f_I \left( {S_{t+1} ,{\cal E}_{t+1} } \right)$, where $f_I $ is the function that infers the network structure based on the relationship between the queried nodes and their neighbors.

\subsection{Problem Formulation}

In this subsection, we formulate a community detection problem in networks where the topology is initially unknown. It is worth noting that the problem of community detection in partially observable networks is much easier to solve and can be thought of as a special case of our problem. This is because, when a certain portion of the network structure has already been explored, this status can be viewed as community detection in partially observable networks. To formulate the problem, we aim to jointly find the community-affiliation embedding matrix {\bf F} and a sequence of node queries $S_T = \left\{ {v_0 ,v_1 , \cdots ,v_{T - 1} } \right\}$ that leads to the finally explored edges ${{\cal E}_T } $, given the query budget $T$, by retrieving information both on node metadata and queried nodes. The likelihood $\mathbb{P}\left( {\mathcal{G}^{\left( T \right)} ,\mathcal{X}\left| \bf F \right.} \right)$ evaluates which affiliation embedding matrix {\bf F} would make the given inferred network $\mathcal{G}^{\left( T \right)} $ and node metadata $\mathcal{X}$ more probable.

If some nodes belong to common communities according to ${\bf F}$, then they are more likely to be connected in $\mathcal{G}^{\left( T \right)}$, and/or to share common metadata from $\mathcal{X}$. This allows us to discover the communities ${\bf F}$ from the network $\mathcal{G}^{\left( T \right)}$ and node metadata $\mathcal{X}$. We formulate this problem by jointly finding the optimal ${\bf F}^*$ and $S_T^ * $ in the sense of maximizing the likelihood $\mathbb{P}\left( {\mathcal{G}^{\left( T \right)} ,\mathcal{X}\left| \bf F \right.} \right)$ as follows:
\begin{equation}
\left( {{\bf F}^ *  ,S_T^* } \right) = \mathop {\arg \max }\limits_{{\bf F} \ge 0,S_T  \subset \mathcal{V}} \mathbb{P}\left( {\mathcal{G}^{\left( T \right)} ,\mathcal{X}\left| {\bf F} \right.} \right),
\end{equation}
where ${\cal G}^{\left( T \right)}  = \left( {\mathcal{V},\mathcal{E}^{\left( T \right)} } \right) $; $\mathcal{E}^{\left( T \right)}  = \mathcal{E}_T  \cup \mathcal{E}_I^{\left( T \right)} $; and $\mathcal{E}_I^{\left( T \right)}  = f_I \left( {S_T ,{\cal E}_T } \right)$.

In our study, we tackle the following two practical challenges. 
\begin{itemize}
\item To the best of our knowledge, there is no prior attempt to detect overlapping communities in networks that lack structural information. All the existing community detection methods are no longer valid in solving the aforementioned problem;
\item The problem of finding a sequence of node queries $S_T$ among all nodes for estimating $\mathcal{E}$ is NP-hard with an exponential complexity in $N$.
\end{itemize} 

\begin{figure*}[htbp]
\begin{center}
    \includegraphics[width=1.0\linewidth]{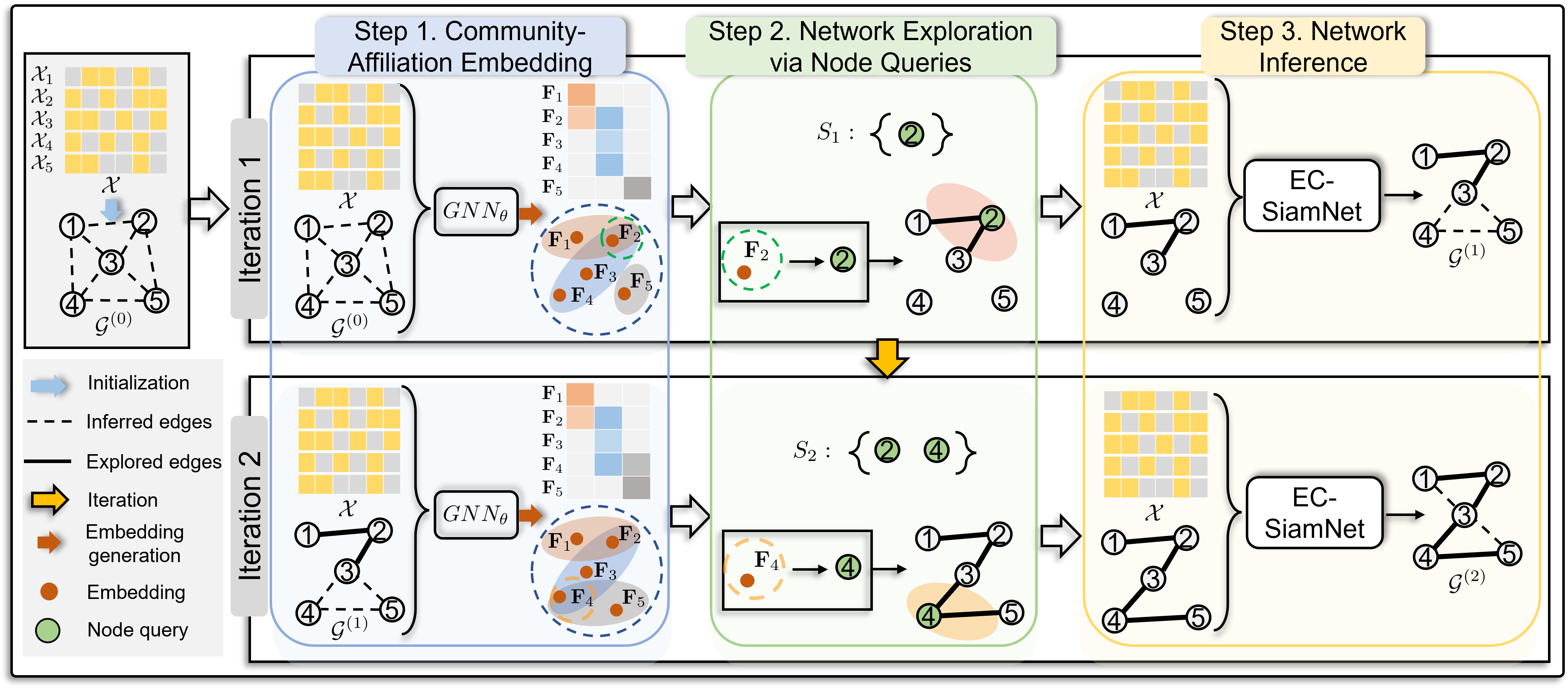}
\end{center}
\captionsetup{skip=-10pt}
   \caption{The schematic overview of \textsf{META-CODE} consisting of three iterative steps: 1) community-affiliation embedding, generated by GNNs to capture both structure–community and metadata–community relationships; 2) network exploration via node queries, which are selected within areas of overlapping communities and distributed across diverse communities; and 3) network inference, which builds potential edges between nodes based on connectivity information from explored edges. Here, the first and second iterations are executed.}
\label{fig:overview}
\vspace{-1.4em}
\end{figure*}

This motivates us to present a scalable yet accurate solution to the problem.

\subsection{Overview of \textsf{META-CODE}}
\label{sec:overview}

In this subsection, we explain our methodology alongside an overview of our \textsf{META-CODE} method. We generate the community-affiliation embedding matrix {\bf F} with the aid of information retrieved from both node metadata and queried nodes. The overall procedure of \textsf{META-CODE} is described in Algorithm \ref{al1}. The algorithm starts with inferring an initial network structure $ \mathcal{G}^{\left( 0 \right)}$ based only on node metadata (\textsf{NetInitilize}---refer to line 2). Then, the algorithm iteratively performs the following three steps within our budget $T$: 1) community-affiliation embedding (\textsf{ComDetect}---refer to line 4); 2) network exploration via node queries (\textsf{QNodeSelect}---refer to line 7); and 3) network inference (\textsf{NetInfer}---refer to line 10). The schematic overview of \textsf{META-CODE} is illustrated in Fig. \ref{fig:overview}.

\begin{algorithm}[htbp]
\begingroup
\DontPrintSemicolon
\KwIn{$\mathcal{V},\mathcal{X}, S_0  = \emptyset, T$}
\KwOut{${\bf F}, S_T$}
\SetKwBlock{Begin}{function}{end function}
\Begin($\textsf{META-CODE}$)
{
  $\mathcal{G}^{(0)} \leftarrow$ \textsf{NetInitilize}($\mathcal{X}$)\;
  
  \For{$t$ {\upshape \bf from} $0$ {\upshape \bf to} $T$}
  {
  ${\bf F} \leftarrow$ \textsf{ComDetect}($\mathcal{G}^{(t)}, \mathcal{X}$)\;

  \If{$t = T$}{\Return{${\bf F},S_T$}}

  $v_t \leftarrow$ \textsf{QNodeSelect}(${\bf F}, S_t$)\;

  $S_{t+1} \leftarrow S_{t}  \cup \left\{ {v_{t} } \right\}$\;
  
  ${\cal E}_{t + 1}  \leftarrow {\cal E}_t  \cup {\cal E}\left( {{\cal N}_{\cal G} \left( {v_{t} } \right),v_{t} } \right)$\;
  
  $\mathcal{E}_I^{\left( t+1 \right)} \leftarrow$ \textsf{NetInfer}$\left( S_{t+1}, \mathcal{E}_{t+1} \right)$\;

  $ \mathcal{G}^{\left( t+1 \right)}  \leftarrow \left( {\mathcal{V} ,\mathcal{E}_{t+1}  \cup \mathcal{E}_I^{\left( t+1 \right)} } \right)$
  }
}
\caption{\textsf{META-CODE}}
\label{al1}
\endgroup
\end{algorithm}

{\bf Step 1: Community-affiliation embedding}. We tackle the challenge of detecting overlapping communities by generating node-level community-affiliation embeddings $\bf F$ through the use of a GNN architecture, which is expressed as:
\begin{equation}
    {\bf{F}}: = GNN_\theta  (\mathcal{G}^{\left( t \right)} ,{\cal X}),
    \label{eq: gnn}
\end{equation}
where $\theta$ is the trainable parameters in the GNN model. Here, if multiple elements are higher than a certain threshold in ${\bf F}_u$, the node belongs to multiple communities.

To ensure that the model accurately captures both structural and attribute information, we train the GNN parameters $\theta$ by establishing our own reconstruction loss in terms of preserving structural and attribute information, which will be specified in Section \ref{sec:step1_detail}. In contrast, prior work \cite{shchur2019overlapping} focused solely on preserving structural information.

{\bf Step 2: Network exploration via node queries}. Based on the obtained community-affiliation embedding {\bf F}, We explore the network through a node query process by selecting a node to be queried in a way that accelerates network exploration, which shall be empirically validated in Section \ref{sec:RQ3}. For example, as depicted in Fig. \ref{fig:overview}, when node $v_2$ is queried in the first iteration, we are capable of discovering its two neighbors, namely nodes $v_1$ and $v_3$.

{\bf Step 3: Network inference}. We adopt the SiamNet model \cite{bromley1993signature}, which has been widely used in computer vision and pattern analyses \cite{chopra2005learning, liu2019exploiting, qi2018hedging}, as the inference function $ f_I \left( {S_{t+1} ,{\cal E}_{t+1} } \right)$ that infers the edges ${\cal E}_I^{\left( t+1 \right)}$. In our study, we propose {\it edge connectivity-based Siamese neural network (EC-SiamNet)} as a potent architecture that learns the similarity between two nodes based on connectivity information obtained by the explored edges. This network inference allows us to fully utilize the explored network information from Step 2, resulting in improved community detection performance.

The three steps in our method are iteratively carried out, feeding the updated inferred network $ \mathcal{G}^{\left( t+1 \right)} $ and node metadata $\mathcal{X}$ as input to the GNN model in each iteration to produce increasingly accurate community-affiliation embeddings ${\bf F}$. This iterative process is terminated when a given budget of node queries $T$ is reached.

\section{Proposed \textsf{META-CODE} method}

In this section, we elaborate on each step in our \textsf{META-CODE} method. We also present the theoretical analyses including the effectiveness of network exploration in \textsf{META-CODE} and the computational complexity of \textsf{META-CODE}.

\subsection{Initial Network Inference}
\label{sec:initial}
Initially, we are faced with a lack of structural information. To overcome this, we need to infer an initial network structure based solely on the available node metadata. 

To this end, as the first step in initial network inference, we adopt the multi-assignment clustering (MAC) approach \cite{streich2009multi}. This involves determining the initial community memberships ${\bf C} \in \left\{ {0,1} \right\}^{N \times K}$, which assigns nodes to multiple communities based on their similarity with community prototypes derived from the node metadata $\mathcal{X}$. Here, ${\bf C}_{uk}$ indicates whether node $u$ belongs to community $k$. The assignment process relies on the decomposition of $\mathcal{X}$ into ${\bf C} \otimes {\bf U} $, where the matrix ${\bf U} \in \left\{ {0,1} \right\}^{K \times D}$ is community prototypes, and ${\bf U}_{k \cdot } $ denotes the prototype of community $k$. Here, the operator $\otimes$ is defined such that the $d$-th attribute of node $u$ satisfies 
\begin{equation}
    {\cal X}_{ud}  =  \vee _{k = 1}^K \left[ {{\bf{C}}_{uk}  \wedge {\bf{U}}_{kd} } \right],
\end{equation}
which signifies that the nodes with the same community memberships tend to share similar node metadata, where $\mathcal{X}_{ud}$ is the $d$-th attribute of node $u$.

Then, as the second step in initial network inference, we infer network ${\cal G}^{\left( 0 \right)}  = \left( {\mathcal{V} ,\mathcal{E}^{\left( 0 \right)} } \right) $ from the initial community memberships ${\bf C}$ using the AGM \cite{yang2012community}, where ${\mathcal{E}^{\left( 0 \right)} } $ represents the set of edges initially inferred. Here, each edge $\left( {u,v} \right) \in \mathcal{E}^{\left( 0 \right)} $ is created with the probability of
\begin{equation}
    p\left( {u,v} \right) = 1 - \prod\nolimits_{i = 1}^{c_{uv} } {\left( {1 - p} \right)},
\label{eq:comm_gen}
\end{equation}
where $p$ is the probability of an edge formed between two nodes within a community and $c_{uv} $ is the number of communities shared by nodes $u$ and $v$ according to the community memberships ${\bf C}$. This implies that, if a pair of nodes $u$ and $v$ belongs to more communities in common, then the probability of creating edge $\left( {u,v} \right)$ is higher.

\subsection{Community-Affiliation Embedding}
\label{sec:step1_detail}
Let us recall that the community-affiliation embeddings {\bf F} can be derived from the GNN model in (\ref{eq: gnn}). Using the inferred network ${\cal G}^{\left( t \right)}$ and node metadata $\mathcal{X}$, we aim to design our GNN model that generates node-level embeddings represented as a non-negative weight affiliation matrix ${\bf{F}} = GNN_\theta  (\mathcal{G}^{\left( t \right)} ,{\cal X}) $. To achieve this goal, we characterize the following two relationships.

\subsubsection{{\bf Structure--Community Relationship}} According to the AGM, if two nodes belong to more communities in common, then they are more likely to be interconnected in the network (refer to (\ref{eq:comm_gen})). This implies that there exists a strong correlation between the network structure and community affiliation. 

To characterize the relationship between the network structure and community affiliation in network ${\cal G}^{\left( t \right)}$, we employ a probabilistic generative model in \cite{yang2013overlapping}. This involves modeling the probability of having an edge between nodes $u$ and $v$ as $1 - \exp \left( { - {\bf F}_u {\bf F}_v^T } \right) $ based on the community-affiliation embedding matrix ${\bf F}$. A higher value of $1 - \exp \left( { - {\bf F}_u {\bf F}_v^T } \right) $ indicates a higher probability of having an edge between nodes $u$ and $v$ in the network, which implies that, if a pair of nodes belongs to more communities in common, then the probability that the node pair is connected is higher. For example, as illustrated in Fig. \ref{fig:relation}, nodes $v_1$ and $v_2$ sharing the same community memberships have a higher probability of being connected compared to nodes that may belong to different communities such as nodes $v_1$ and $v_3$.

In this context, to determine which community-affiliation embeddings ${\bf F}$ would make the given inferred network $\mathcal{G}^{(t)}$ more probable, we formulate our loss function $\mathcal{L}_1 ({\bf F})$ as:
\begin{equation}
\begin{array}{l}
 \mathcal{L}_1 \left( {\bf{F}} \right) =  - \log \mathbb{P} \left( {\mathcal{G}^{\left( t \right)} |{\bf{F}}} \right) \\ 
  \;\;\;\;\;\;\;\;\;\;\;= -\!\!\!\!\! \sum\limits_{\left( {u,v} \right) \in \mathcal{E}^{\left( t \right)} }  \log \left( {1 - \exp \left( { - {\bf{F}}_u {\bf{F}}_v^T } \right)} \right) +\!\!\!\! \sum\limits_{\left( {u,v} \right) \notin \mathcal{E}^{\left( t \right)} }\!\!\! {{\bf{F}}_u {\bf{F}}_v^T },  \\ 
 \end{array}
\label{eq: structue_community}
\end{equation}
where $\mathcal{E}^{(t)}$ is the set of edges in $\mathcal{G}^{(t)}$.

\begin{figure}[t]
\begin{center}
    \includegraphics[width=1.0\linewidth]{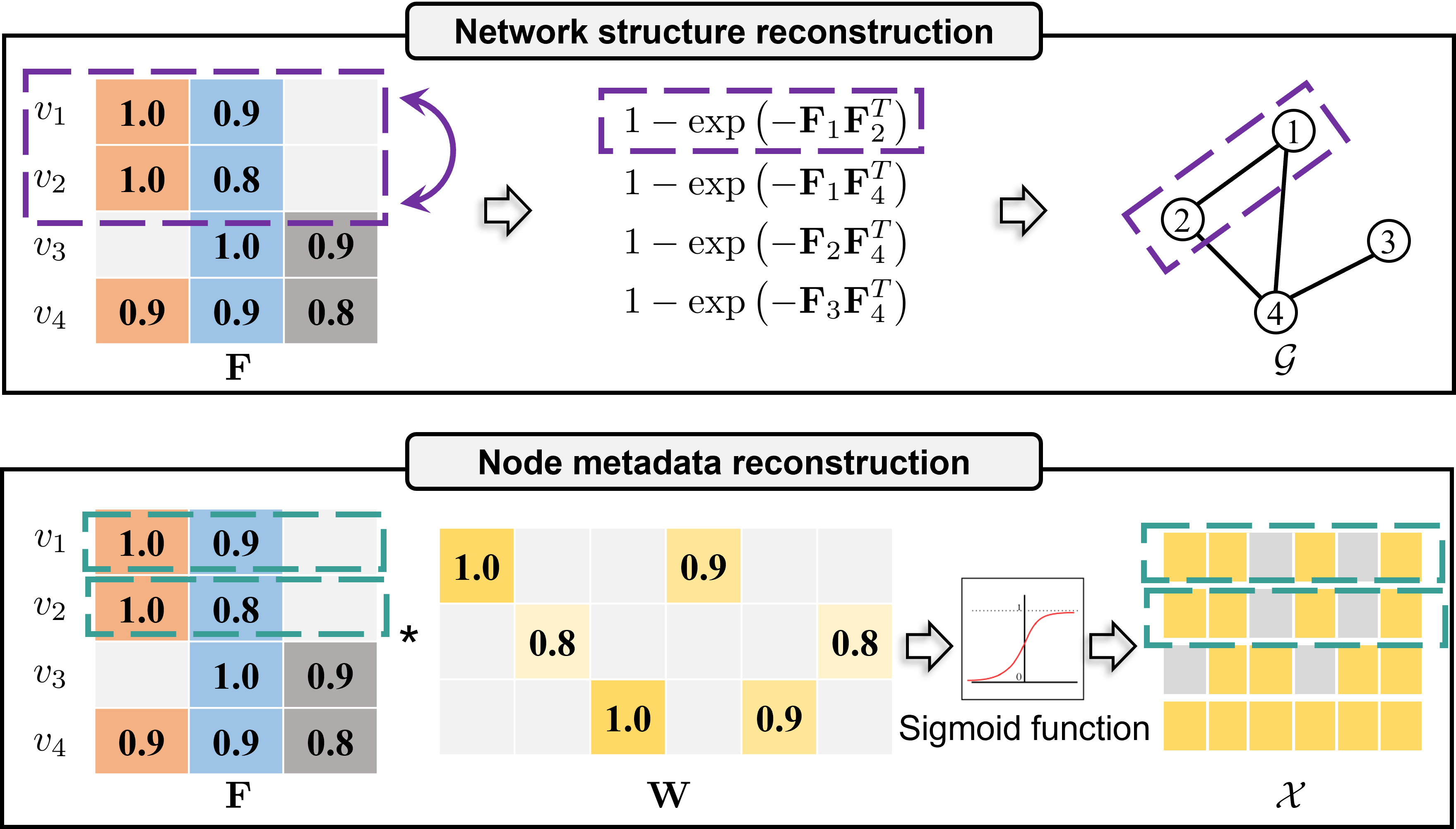}
\end{center}
\captionsetup{skip=-5pt}
   \caption{An example illustrating how network structure $\mathcal{G}$ and node metadata $\mathcal{X}$ can be reconstructed from the community-affiliation embedding matrix ${\bf F}$.}
\label{fig:relation}
\vspace{-1.5em}
\end{figure}

\subsubsection{{\bf Metadata--Community Relationship}} Intuitively, if nodes are in the same community, then they are likely to share common node attributes \cite{yang2013community}. As depicted in Fig. \ref{fig:relation}, nodes $v_1$ and $v_2$ with the same community membership share common node metadata. We next state how to characterize the relationship between the node metadata and community structure, which is built upon the attribute modeling in \cite{yang2013community} by hypothesizing that incorporating a node's community memberships can facilitate the prediction of each attribute associated with the node.

More formally, by adopting a logistic model to build the relationship, we can probabilistically model the $d$-th attribute of node $u\in\mathcal{V}$ in the binary-valued node metadata $\mathcal{X}$, denoted by $\mathcal{X}_{ud}$, as the sigmoid function:
\begin{equation}
    {\cal Q}_{ud}  = \frac{1}{{1 + \exp \left( { - \sum\nolimits_k {{\bf F}_{uk} {\bf W}_{dk}} } \right)}},
\label{eq: metadata_community}
\end{equation}
where ${\bf{F}}_{uk} $ is the $k$-th entry in ${\bf{F}}_{u} $ and ${\bf W}_{dk} $ is the $\left( {d,k} \right) $-th entry of weight matrix ${\bf W} \in \mathbb{R}^{D \times K} $, indicating the relevance of community $k$ to the $d$-th node attribute. As illustrated in Fig. \ref{fig:relation}, the corresponding node metadata can be reconstructed from $\mathcal{Q}_{ud}$. In this context, we formulate our loss function $\mathcal{L}_2 ({\bf F})$ as:
\begin{equation}
\begin{array}{l}
 \mathcal{L}_2 \left( {\bf{F}} \right) =  - \log \mathbb{P} \left( {\mathcal{X}|{\bf{F}},{\bf{W}}} \right) \\ 
 \:\:\:\:\:\:\:\:\:\:\:\:\:\:= - \sum\limits_{u,d} {\left( {{\cal X}_{ud} \log {\cal Q}_{ud}  + \left( {1 - {\cal X}_{ud} } \right)\log \left( {1 - {\cal Q}_{ud} } \right)} \right)}.  \\ 
 \end{array}
 \label{eq: metadata_reconstruction}
\end{equation}

\subsubsection{Model Training}
Based on the two aforementioned relationships, we establish our reconstruction loss as follows:
\begin{equation}
    \begin{aligned}
    \mathcal{L}\left ( {\bf F} \right )&=\mathcal{L}_1({\bf F})+\eta \mathcal{L}_2({\bf F}) \\
      &= - \sum\limits_{\left( {u,v} \right) \in {\cal E}^{\left( t \right)} } \!\! {\log \left( {1 - \exp \left( { - {\bf F}_u {\bf F}_v^T } \right)} \right) +\!\!\!\! \sum\limits_{\left( {u,v} \right) \notin {\cal E}^{\left( t \right)} } {{\bf F}_u {\bf F}_v^T } } \\
    \end{aligned}
\notag
\end{equation}
\begin{equation}
    \;\;\;\;\;\;\;- \eta \sum\limits_{u,d} {\left( {\mathcal{X}_{ud} \log \mathcal{Q}_{ud}  + \left( {1 - \mathcal{X}_{ud} } \right)\log \left( {1 - \mathcal{Q}_{ud} } \right)} \right)},
     \label{eq:loss}
\end{equation}
where $\eta  \ge 0 $ is the hyperparameter to balance between two terms. Note that the first and second terms in (\ref{eq:loss}) aim at reconstructing the network structure and node metadata, respectively. We update the parameters $\theta $ of our GNN model in the sense of minimizing the loss function. 

During training, if node $u$ belongs to multiple communities, then the trainable parameter $\theta $ is optimized in such a way of increasing the values of belonging elements in ${\bf F}_u$ for $u$ ({\bf structure-community relationship}). Additionally, $\theta $ is also optimized in such a way of making ${\bf F}_u$ and ${\bf F}_v$ similar if nodes $u$ and $v$ share common node attributes ({\bf metadata-community relationship}). After training, overlapping communities are assigned to each node based on the final community-affiliation embeddings ${\bf F}$, with memberships determined by elements that exceed a certain threshold.

\vspace{-0.3em}

\subsection{Network Exploration via Node Queries}
\label{sec:step2_detail}
\begin{figure} 
    \begin{center}
        \includegraphics[width=1.0\linewidth]{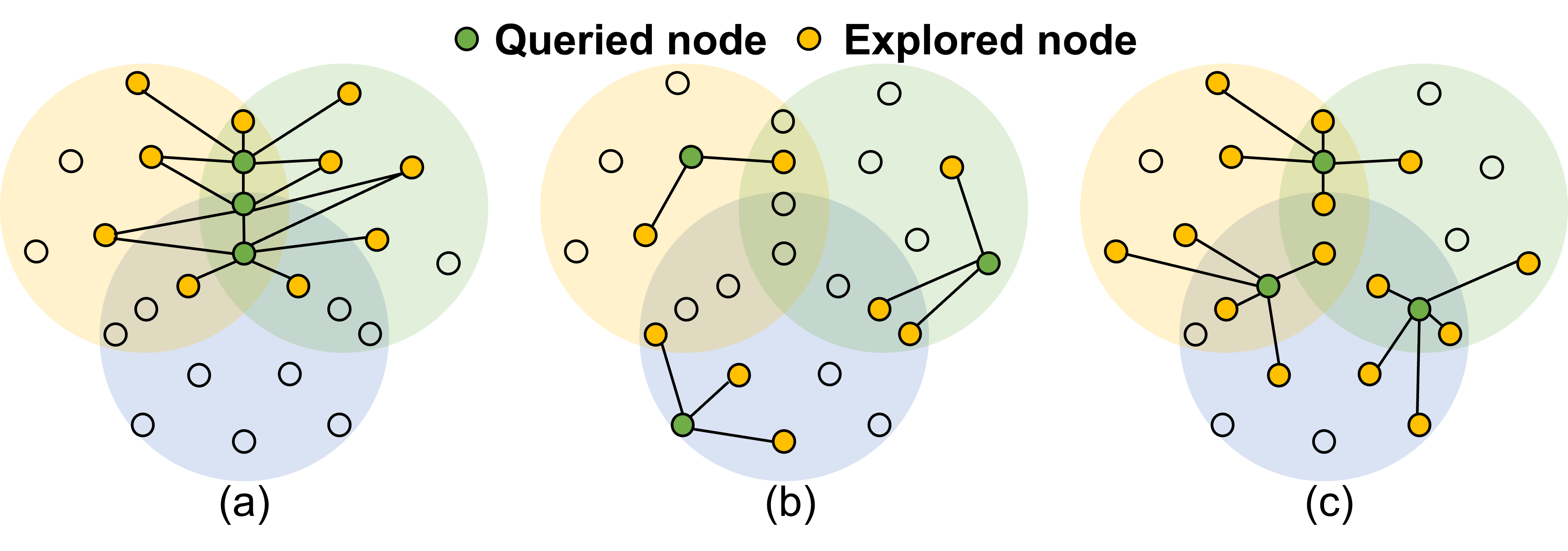}
    \end{center}
    \captionsetup{skip=-10pt}
    \caption{Network exploration with different strategies for query node selection when the underlying true network has three overlapping communities. (a) Selection of query nodes with the highest degree. (b) Selection of query nodes in non-overlapping regions. (c) Selection of query nodes that belong to multiple communities and are distributed across diverse communities.}
    \label{fig:network_ex} 
    \vspace{-1.7em}
\end{figure}

As stated in Section \ref{sec:overview}, we aim to maximize the number of discovered neighbors for each queried node to support network exploration, utilizing the updated community-affiliation embedding {\bf F}. To this end, we are interested in how to select node queries $S_t$. Although querying high-degree nodes seems ideal for fast network exploration, it may not always be true in the overlapping community detection task. A simple query strategy does not necessarily result in fast network exploration; for example, in Figs. \ref{fig:network_ex}a and \ref{fig:network_ex}b, selecting high-degree nodes (i.e., those with the degree of 6) or nodes in non-overlapping regions (i.e., those belonging to only one community) may result in a limited number of explored neighbors. Moreover, such topology-aware query strategies would not be possible due to the {\it neighbor unawareness} of nodes beforehand. 

In our study, the node query strategy is built upon our claim that each queried node should not only belong to {\it more communities} but also be distributed over {\it diverse communities} for faster network exploration. As shown in Fig. \ref{fig:network_ex}c, by selecting nodes that are (i) within areas of {\it overlapping} communities and (ii) distributed across {\it diverse} communities, we are capable of exploring different regions within each community, thus expediting the expansion of the explored network. This eventually leads to a higher accuracy of community detection, which will be empirically validated in Section \ref{sec:RQ3}.

In this context, we select the $(t+1)$-th node to query for $t\in\{0,\cdots,T-1\}$ as follows:
\begin{equation}
    v_t = \mathop {\arg \max }\limits_v \left( {\left\| {{\bf F}_v } \right\|_1  + \lambda\left( {1 - \frac{1}{{t+1}}\sum\limits_{u \in S_t} {\text{sim}\left( {{\bf F}_v ,{\bf F}_u } \right)} } \right)} \right),
\label{eq:strategy}
\end{equation}
where $\left\|  \cdot  \right\|_1$ represents $L_1$-norm of a vector, $\lambda \ge 0$ is the hyperparameter that balances between the two terms, and $\text{sim}\left( { \cdot , \cdot } \right)$ is the cosine similarity. Here, the first term in (\ref{eq:strategy}) maximizes community affiliation strength, prioritizing nodes with strong associations to multiple communities to facilitate faster coverage across the network; and the second term in (\ref{eq:strategy}) promotes community diversity, encouraging the selection of nodes from different communities than those previously queried, ensuring balanced exploration and avoiding over-concentration in any single community.

Once we have selected the queried node $v_t$, it is added to the sequence $S_{t+1}  = S_{t}  \cup \left\{ {v_{t} } \right\} $. Then, we can discover its neighbors $\mathcal{N}_\mathcal{G}(v_t)$ and expand the observable subnetwork by updating the set of explored edges ${\cal E}_{t+1}  = {\cal E}_{t}  \cup {\cal E}\left( {{\cal N}_{\cal G} \left( {v_{t} } \right),v_{t} } \right)$.

\subsection{Network Inference}
\label{sec:step3_detail}
During the query process, the set of explored edges, ${\cal E}_{t+1}$, increases as a result of the node queries $S_{t+1}$. This newly acquired information, including the queried nodes and explored edges, can then be fully utilized to generate potential edges within the network. By leveraging this expanded knowledge of the network structure, we can enhance the model’s capability to infer likely connections among nodes, ultimately supporting a more comprehensive network structure and improving community detection \cite{tran2024meta}.

\begin{figure}[t]
\begin{center}
    \includegraphics[width=1.0\linewidth]{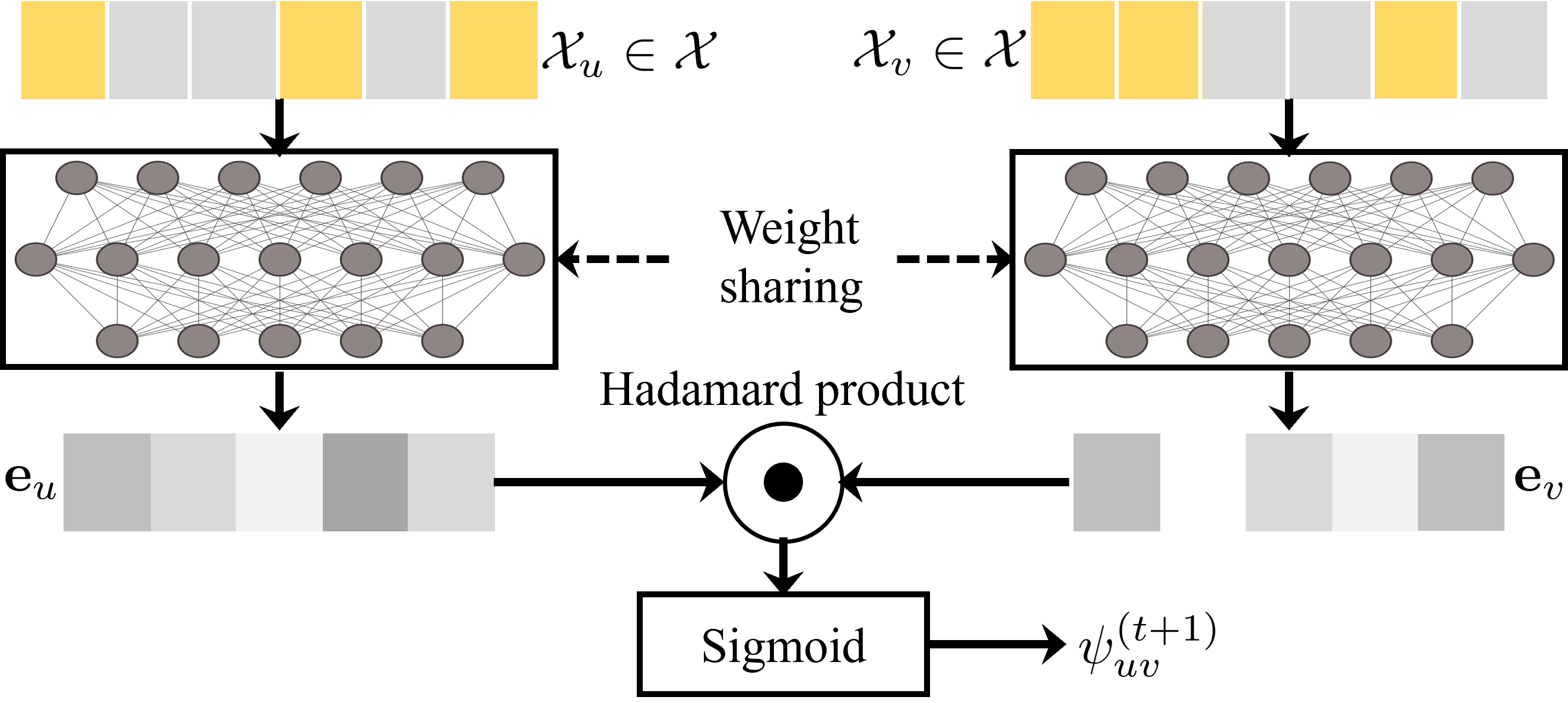}
\end{center}
\captionsetup{skip=-5pt}
   \caption{The architecture of our EC-SiamNet for network inference.}
\label{fig:siamese}
\vspace{-1.0em}
\end{figure}

To do this, the inference function $ f_I \left( {S_{t+1} ,{\cal E}_{t+1} } \right)$ is designed by using our EC-SiamNet model, which incorporates the network structure and node metadata to infer edges over the network, whereas only the node metadata $\mathcal{X}$ are utilized in the initial network inference step (see Section \ref{sec:initial}). We now explain how to train EC-SiamNet and how to compute the similarity score $\Psi ^{\left( t+1 \right)} $ in Step 3 of Section \ref{sec:overview}. In our study, we take advantage of the ``homophily principle'' in social networks, which refers to the tendency of an individual node to connect with similar nodes \cite{tang2013exploiting}, and is used to learn embeddings that precisely capture the homophily effect. A trained EC-SiamNet inductively maps new input data (i.e., unexplored nodes) whose connectivity information to the training data (i.e., explored nodes) is unknown. As shown in Fig. \ref{fig:siamese}, our EC-SiamNet architecture comprises twin multilayer perceptron networks (MLPs) that take two node metadata $\mathcal{X}_u$ and $\mathcal{X}_v$, corresponding to nodes $u$ and $v$, respectively, as input. The MLPs act as encoders that generate embedding vectors, denoted as ${{\bf e}_u }$ and ${{\bf e}_v }$ for nodes $u$ and $v$, respectively. Similarly as in \cite{wang2017predictive}, we compute the Hadamard product of two embedding vectors ${\bf e}_u$ and ${\bf e}_v$ and then utilize a sigmoid output layer to infer $\psi _{uv} ^{\left( t+1 \right)} \in \Psi^{(t+1)}$ from the product. That is, we compute
\begin{equation}
    \psi _{uv} ^{\left( t+1 \right)}  = \text{sigmoid}\left( {{\bf e}_u  \odot {\bf e}_v } \right),
\end{equation}
where $\odot $ denotes the Hadamard product.

Next, we optimize the model parameters ${\bf W}_{Sia} $ of EC-SiamNet. To this end, we establish a contrastive loss built upon the approach in \cite{hadsell2006dimensionality}. Specifically, for each mini-batch used in stochastic gradient descent, we sample a pair of nodes $u,v \in \mathcal{V} $ whose connection is certain according to $\mathcal{E}_{t+1}$. Here, to utilize the sparsity of real-world networks and balance between the sampled positive and negative node pairs, we apply negative sampling \cite{armandpour2019robust}. Then, our goal is to find parameters ${\bf W}_{Sia} $ in the sense of minimizing the following loss function:
\begin{equation}
    \begin{array}{l}
         \mathcal{L}\left( {{\bf W}_{Sia} ,\mathcal{X}_v ,\mathcal{X}_v } \right) = \left( {1 - I_{uv} } \right)\frac{1}{2}\left[ {\text{sim}\left( {{\bf e}_u ,{\bf e}_v } \right)} \right]^2  \\ 
         \;\;\;\;\;\;\;\;\;\;\;\;\;\;\;\;\;\;\;\;\;\;\;\;\;\;\;\;\;\; + I_{uv} \frac{1}{2}\left[ {\max \left( {0,r - \text{sim}\left( {{\bf e}_u ,{\bf e}_v } \right)} \right)} \right]^2, \\ 
    \end{array}
    \label{eq:loss_siamese}
\end{equation}
where $r$ is the so-called “margin” hyperparameter \cite{hadsell2006dimensionality}; and $I_{uv} = 1$ if $\left( {u,v} \right) \in \mathcal{E}_{t+1} $ and $I_{uv} = 0$ otherwise. Note that we make use of information of the explored edges (i.e., the parameter $I_{uv}$) in the stage of EC-SiamNet optimization.
    
The motivation behind the loss function in (\ref{eq:loss_siamese}) is two-fold: 1) to create an attractive force between the embeddings of node pairs with similar metadata, and 2) to generate a repulsive force between dissimilar node pairs that pushes them apart until the distance between them in the embedding space exceeds a certain radius $r$.

By calculating the similarity score $\Psi^{(t+1)}$ via the proposed EC-SiamNet model, we are able to more precisely obtain the set of inferred edges $\mathcal{E}_I^{\left( t+1 \right)} $, which is then incorporated into the edges $\mathcal{E}^{\left( t+1 \right)} $ of network ${\cal G}^{\left( t+1 \right)}$, that is, $\mathcal{E}^{\left( t+1 \right)}  = \mathcal{E}_{t+1}  \cup \mathcal{E}_I^{\left( t+1 \right)} $. The updated network ${\cal G}^{\left( t+1 \right)}$ is then used as the input to the GNN model in Step 1 of the next iteration to obtain an enhanced representation of community-affiliation embedding matrix {\bf F}. This procedure is repeated until the $T$-th node query is reached.

\begin{remark}
\label{remark:con}
    Let us address how EC-SiamNet converges over iterations. It is worth noting that, as the query process continues, we are capable of having access to more node pairs for training from the explored network. As far as the homophily principle holds, adopting our EC-SiamNet model enables us to infer edges more accurately, leading to the convergence of the network ${\cal G}^{\left( t \right)}$ to its underlying true network $\mathcal{G}$. This results in more precise community-affiliation embeddings.
\end{remark}

\subsection{Theoretical Analysis}
\label{sec:theorem}
In this subsection, we theoretically analyze the effectiveness of network exploration via node queries in \textsf{META-CODE}. We also theoretically show the computational complexity of \textsf{META-CODE}. 

\subsubsection{Theoretical Findings in \textsf{META-CODE}}
We first analytically show the effectiveness of our node query strategy. For ease of analysis, we make three assumptions for an underlying network $\mathcal{G}=(\mathcal{V},\mathcal{E})$ as follows:

\begin{enumerate}
    \item The underlying network $\mathcal{G}$ follows the AGM, where the probability of forming an edge in each community is identically given by $p$.
    \item The underlying network $\mathcal{G}$ is composed of $K\left( { \ge 2} \right)$ communities, having a constrained size difference such that
    \begin{equation}
        N_{\max }  - N_{\min }  \le \varepsilon ,
        \label{eq:assumption2}
    \end{equation}
    where $N_{\max }$ and $N_{\min }$ are the maximum and minimum numbers of nodes in each community, respectively, and $\varepsilon>0$ is an arbitrarily small constant. 
    \item We partition the nodes into $K$ subsets according to the number of communities to which they belong. Let $N_1,N_2,\cdots,N_K$ denote the number of nodes in each subset. Then,
    \begin{equation}
        \left\{ \begin{array}{l}
        \frac{{N_1 }}{N} \le \frac{2}{3} \\
         \mathbb{E}_v \left[ {\mathcal{D}\left| {A_1 } \right.} \right]N_1  \ge \sum\limits_{i = 2}^K {\mathbb{E}_v \left[ {\mathcal{D}\left| {A_i } \right.} \right]N_i }, \\ 
         \end{array} \right.
         \label{eq:assumption3}
    \end{equation}
    where $\mathcal{D}$ is the degree distribution of nodes and $A_i$ is the event such that a certain node $v$ belongs to $i$ communities.

\end{enumerate}
Assumptions 1 and 2 are made to maintain a consistent density within each community for simplicity. Importantly, these assumptions do not fundamentally change the precondition (of theorems to be shown below) that a node belongs to overlapping communities. While relaxing these assumptions would make the proof more complex, the conclusions would remain consistent, as such modifications would not affect the core precondition of a node's membership in overlapping communities. Assumption 3 enforces nodes not to belong {\it overly} to either only a single community or multiple ($\ge2$) communities. Note that many real-world networks (e.g., Ego-Facebook \cite{mcauley2014discovering} and citation networks \cite{shchur2019overlapping}) tend to almost follow Assumption 3. Now, we are ready to show the following two theorems.

\begin{thm} 
\textit{For any nodes $u$ and $v$ belonging to $M$ and $M'$ communities, respectively, in an underlying network $\mathcal{G}=(\mathcal{V},\mathcal{E})$, where $M > M'$, if $\varepsilon  \le \frac{{N_{\min }  - 1}}{K} - 1$, then the following inequality holds:}
\begin{equation}
    \mathbb{E}_{u } \left[ {\mathcal{D}_M } \right] \ge \mathbb{E}_{v } \left[ {\mathcal{D}_{M'} } \right],
\end{equation}
\textit{where $\mathbb{E}_{u } \left[ {\mathcal{D}_M } \right]$ and $\mathbb{E}_{v } \left[ {\mathcal{D}_{M'} } \right]$ are the expectations of degree distributions over the nodes belonging to $M$ and $M'$ communities, respectively.}
\end{thm}

\begin{proof}
    We refer to Appendix A for the proof of this theorem.
\end{proof}

From Theorem 1, one can see that selecting a node belonging to more communities is expected to result in faster network exploration due to the fact that more neighbors of the queried node can be discovered. In other words, nodes within areas of more overlapping communities have higher chances of being connected to other nodes, enabling efficient network exploration, which supports our design principle in query node selection. This is empirically validated in Section V-E2, where selected nodes within areas of more overlapping communities with our node query strategy tends to explore more neighboring neighbors.

\begin{thm}
\textit{For any node $u$ belonging to $M$ $(\ge 2)$ communities in an underlying network $\mathcal{G}=(\mathcal{V},\mathcal{E})$ and any arbitrary node $v$ in $\mathcal{G}$, if $\varepsilon  \le \frac{{N_{\min }  - 1}}{K} - 1$, then the following inequality holds:}
\begin{equation}
    \mathbb{E}_u \left[ {\mathcal{D}_M } \right] \ge \mathbb{E}_v \left[ \mathcal{D} \right],
\end{equation}
\textit{where $\mathbb{E}_{u } \left[ {\mathcal{D}_M } \right]$ is the expectation of degree distributions over the nodes belonging to $M$ communities, while $\mathbb{E}_{v } \left[ {\mathcal{D} } \right]$ is the expectation of degree distributions over all the nodes in $\mathcal{G}$.}
\end{thm}

\begin{proof}
    We refer to Appendix B for the proof of this theorem.
\end{proof}

Theorem 2 further confirms that the proposed node query strategy yields faster network exploration compared to the random selection for node query. Our strategy attempts to select the nodes within areas of overlapping communities, which exhibit denser connections, thus accelerating the expansion of the explored network. This is also empirically validated in Section V-E2, where our node query strategy explores networks more efficiently than random selection.

\subsubsection{Complexity Analysis}
To validate the scalability of our \textsf{META-CODE} method, we analytically show its computational complexity by establishing the following theorem.

\begin{thm}
\textit{The computational complexity of the proposed
\textsf{META-CODE} method is given by $\mathcal{O}\left( {\left| \mathcal{E} \right|} \right)$.}
\end{thm}

\begin{proof}
    We refer to Appendix C for the proof of this theorem.
\end{proof}

From Theorem 3, one can see that the computational complexity of \textsf{META-CODE} scales {\it linearly} with the number of edges in the underlying network. This is empirically validated in Section V-E8.

\section{Experiment Evaluation}

In this section, we describe the datasets used in our evaluation and six benchmark overlapping community detection methods, including a simple variant of \textsf{META-CODE}, for comparison. After describing the performance metrics and experimental settings, we comprehensively evaluate the performance of our \textsf{META-CODE} method and the six benchmark methods.
We design our extensive empirical study to answer the following nine key research questions (RQs):

\begin{itemize}
    \item {\bf RQ1}: How much does \textsf{META-CODE} improve the performance of community detection over benchmark overlapping community detection methods?

    \item {\bf RQ2}: How effective is our query node strategy for fast network exploration?

    \item {\bf RQ3}: How close is the inferred network to the underlying true network?

    \item {\bf RQ4}: How do different GNN backbones influence community detection accuracy?

    \item {\bf RQ5}: How does the performance of overlapping community detection methods behave with respect to modularity with no ground truth communities?

    \item {\bf RQ6}: How does each module in \textsf{META-CODE} contribute to community detection accuracy?

    \item {\bf RQ7}: How do key parameters affect the performance of \textsf{META-CODE}?
    
    \item {\bf RQ8}: How scalable is \textsf{META-CODE} with the size of the network?
    
    \item {\bf RQ9}: What does the visualization of the community detection results produced by \textsf{META-CODE} reveal?
\end{itemize}

\subsection{Datasets}

We conduct our experiments on four real-world datasets across several domains, which include one social networks, Ego-Facebook \cite{mcauley2014discovering}\footnote{As there are a number of ego-networks sampled from the original network, we use Facebook 348 in our experiments. For brevity, we omit the index.}, and three large co-authorship networks, Coauthorship-Engineering (Engineering), Coauthorship-Computer Science (Computer Science), and Coauthorship-Medicine (Medicine) \cite{shchur2019overlapping}. For the sake of consistency, we assume all networks to be undirected. Table \ref{table:datasets} provides a summary of the statistics for each dataset, including the number of nodes, edges, communities, and node features.\footnote{Although there are other social and large co-authorship networks, we have not adopted them since they exhibit similar tendencies to those datasets under consideration.}

\begin{table}[t]
\footnotesize
\centering
\captionsetup{skip=0pt}
  \caption{The statistics of the datasets.}
  \begin{tabular}{cccccl}
    \toprule
    Dataset& Nodes & Edges & Communities & Features\\
    \midrule
    Facebook & 227 & 3,192 & 14 & 161  \\
    Engineering& 14,927& 49,305 & 16 & 4,839 \\ 
    Computer Science & 21,957 & 96,750 & 18 & 7,793\\
    Medicine & 63,282 & 810,314 & 17 & 5,538\\
  \bottomrule
\end{tabular}
\label{table:datasets}
\vspace{-2.0em}
\end{table}

\subsection{Benchmark Methods}

Since there is {\it no prior work} on overlapping community detection in networks with unknown topology, in our experiments, we consider five competitive benchmark methods for overlapping community detection, including one NMF-based community detection method (BIGCLAM \cite{yang2013overlapping}), one generative model-based community detection methods (vGraph \cite{sun2019vgraph}), and three GNN-based community detection methods (DMoN \cite{tsitsulin2023graph}, CLARE \cite{wu2022clare}, and NOCD \cite{shchur2019overlapping}). Note that these methods were originally developed for {\it fully observable networks}, which require an extra step to infer the network structure. To this end, we employ two sampling strategies for network exploration via node queries: random sampling (RS) and depth-first search (DFS).\footnote{Although other sophisticated sampling strategies such as biased random walks can also be employed, they have not been employed in our study since they have similar tendencies to those of RS and DFS.} The benchmark methods are implemented similarly as in \textsf{META-CODE}, with iterative community detection performed alongside the explored network through sampling given a budget of node queries. We implement all these benchmark methods using the parameter settings described in their original articles. 

Note that, CLARE, NOCD, and DMoN make use of both the network structure and the node metadata for detecting communities. Additionally, in our experiments, we take into account our own simple variant of \textsf{META-CODE}, dubbed \textsf{META-CODE}{\tiny sim}, developed for networks with unknown topology. This baseline method starts with randomly querying a given percentage of nodes and then discovering the neighbors of these queried nodes for network exploration. As the next step, EC-SiamNet in Section \ref{sec:step3_detail} is utilized to infer edges based on the explored network. Finally, the GNN model in Section \ref{sec:step1_detail} is employed to generate the community-affiliation embeddings ${\bf F}$. Note that, unlike the original \textsf{META-CODE}, \textsf{META-CODE}{\tiny sim} runs {\it only once} without iterations.

\subsection{Performance Metrics}

To evaluate the performance of community detection, we quantify the degree of agreement between the ground truth communities and the detected communities by adopting the NMI \cite{mcdaid2011normalized} and the average $\text{F}_1$ score (Avg$\text{F}_1$) \cite{yang2013overlapping}. Note that both metrics have values within the range of $\left[ {0,1} \right] $, with higher values indicating better performance. The availability of ground truth communities enables us to quantitatively evaluate the performance of community detection methods using metrics that measure the level of correspondence between the detected and ground truth communities. We consider a set of ground truth communities $\mathcal{C} $ and a set of detected communities $\hat{\mathcal{C}} $. To evaluate the performance of community detection by quantifying the degree of agreement between the ground truth communities and the detected communities, we adopt the following two performance metrics:
\begin{itemize}
    \item {\bf Normalized Mutual Information (NMI)} \cite{mcdaid2011normalized}. Assume that the community assignments are $x_i$ and $y_i$, where $x_i$ and $y_i$ indicate the labels of node $i$ in the ground truth communities $\mathcal{C} $ and the detected communities $\hat{\mathcal{C}} $, respectively. When the labels $x$ and $y$ are the values of two random variables $X$ and $Y$ following a joint distribution $\mathbb{P}\left( {x,y} \right) = \mathbb{P}\left( {X = x,Y = y} \right) $, the NMI between $\mathcal{C} $ and $\hat{\mathcal{C}} $ is given by
    \[
    NMI\left( {\mathcal{C}  ,\hat{\mathcal{C}}} \right) = 1 - \frac{1}{2}\left( {\frac{{H\left( {X\left| Y \right.} \right)}}{{H\left( X \right)}} + \frac{{H\left( {Y\left| X \right.} \right)}}{{H\left( Y \right)}}} \right),
    \]
    where $H\left( X \right) \!\!=\!\!  - \sum\nolimits_x {\mathbb{P}\left( X \right)\log \mathbb{P}\left( X \right)} $ is the Shannon entropy of $X$ and $H\left( {X\left| Y \right.}\! \right)\!\! =\!\!  - \sum\nolimits_{x,y} {\mathbb{P}\left( {x,y} \right)\log \mathbb{P}\left( {x\left| y \right.} \right)} $ is the conditional entropy of $X$ given $Y$.

    \item {\bf Average ${\bf F}_1$ score (Avg${\bf F}_1$)} \cite{yang2013overlapping}. To compute the ${\rm F}_1$ score, which is the harmonic mean of Precision and Recall, we need to determine which $C_i  \in \mathcal{C} $ corresponds to which $\hat C_i  \in \hat{\mathcal{C}} $. The Avg${\rm F}_1$ is defined as the average of the ${\rm F}_1$ score of the best-matching ground truth community to each detected community and the ${\rm F}_1$ score of the best-matching detected community to each ground truth community and is expressed as
    \[
        \begin{array}{l}
         {\rm{AvgF}}_1 \left( {{\cal C},\widehat{\cal C}} \right) = \frac{1}{2}\left( {\frac{1}{{\left| {\cal C} \right|}}\sum\limits_{C_i  \in {\cal C}} {{\rm{F}}_1 \left( {C_i ,\hat C_{g\left( i \right)} } \right)} } \right. \\ 
         \;\;\;\;\;\;\;\;\;\;\;\;\;\;\;\;\;\;\;\;\;\;\left. { + \frac{1}{{\left| {\widehat{\cal C}} \right|}}\sum\limits_{\hat C_i  \in \widehat{\cal C}} {{\rm{F}}_1 \left( {C_{g'\left( i \right)} ,\hat C_i } \right)} } \right), \\ 
         \end{array}
    \]
    where $g\left( i \right) = \mathop {\arg \max }\limits_j {\rm F}_1\left( {C_i ,\hat C_j } \right) $ and $g'\left( i \right) = \mathop {\arg \max }\limits_j {\rm F}_1\left( {C_j ,\hat C_i } \right) $.

\end{itemize}

\subsection{Experiment Setup}

As a default experiment setting, we conduct experiments for \textsf{META-CODE} with a fixed query budget $T$ as a percentage of all nodes in the underlying network, varying from 10\% to 40\% with an increment of 10\% across all datasets, as shown in Table \ref{tab:comparison}.\footnote{We again state that, when a certain portion of the network structure has been explored, this status can be viewed as community detection in partially observable networks.} We adopt GCN \cite{kipf2016semi} with 2 layers to extract the community-affiliation embeddings and two-layer MLPs as the backbone of our EC-SiamNet for network inference. Each experiment is conducted 5 times to evaluate the average performance. The final community assignments for each node $u$ are determined by the elements in ${\bf F}_u$ that exceed a certain threshold of 0.5. All experiments are carried out with Intel (R) 12-Core (TM) E5-1650 v4 CPUs @ 3.60 GHz and 64GB RAM.

It is noted that, in the co-authorship datasets (i.e., Engineering and Computer Science), we perform network inference on every 1\% of query nodes, instead of every iteration, for computational efficiency. This is due to the fact that obtaining information from only a single queried node's neighbors does not lead to a significant improvement in network inference for such large datasets. We describe the experiment settings of neural networks (i.e., the GNN and MLP models) in our \textsf{META-CODE} method. We adopt GCN \cite{kipf2016semi} with 2 layers to extract the community-affiliation embeddings and two-layer MLPs as the backbone of our EC-SiamNet for network inference. PyTorch Geometric \cite{fey2019fast}, a geometric deep learning extension library in PyTorch, is used to implement the GCN and MLPs. We set the hidden layer dimensions of each GCN and MLP to 128 and 256, respectively. As in plenty of studies (see \cite{yang2012community, yang2013overlapping, yang2013community, shchur2019overlapping, tran2021community} and references therein), the number of communities is assumed to be available. The Adam optimizer \cite{kingma2014adam} with a learning rate of 0.001 and 0.05 is used to train the GNN and MLPs, respectively. Moreover, we tune the hyperparameters $\eta$ and $\lambda$ within designated ranges of $\eta  \in \left[ {1.0,2.0} \right] $ and $\lambda  \in \left[ {1.0,4.0} \right] $.

\subsection{Results and Analyses}

\begin{table*}[!t]\centering
\setlength\tabcolsep{5.0pt}
\scriptsize
\captionsetup{skip=0pt}
  \caption{Performance comparison of \textsf{META-CODE}, \textsf{META-CODE}{\tiny sim}, and 10 competing methods in terms of two metrics (average $\pm$ standard deviation) when different percentages (\%) of nodes are queried among $N$ nodes. Here, the best and second-best performers are highlighted by bold and underline, respectively.}
  \label{tab:comparison}
  \begin{tabular}{cc|cc|cc|cc|ccl}
    \toprule[1pt]
    \multicolumn{2}{c|}{}&\multicolumn{2}{|c|}{10\% queried nodes}&\multicolumn{2}{c|}{20\% queried nodes}&\multicolumn{2}{c|}{30\% queried nodes}&\multicolumn{2}{c}{40\% queried nodes}\\
    \cmidrule{1-10}
           Dataset&Method & NMI& Avg$\text{F}_1$ & NMI& Avg$\text{F}_1$ & NMI& Avg$\text{F}_1$& NMI& Avg$\text{F}_1$\\
    \midrule[1pt]
    \multirow{11}*{\rotatebox{90}{Facebook}} &
    BIGCLAM+RS & 0.0263\tiny$\pm$0.0056& 0.4349\tiny$\pm$0.0213& 0.0539\tiny$\pm$0.0193& 0.5025\tiny$\pm$0.0681& 0.0703\tiny$\pm$0.0135& 0.5896\tiny$\pm$0.0117& 0.1083\tiny$\pm$0.0162& 0.6503\tiny$\pm$0.1299\\
    & BIGCLAM+DFS& 0.0379\tiny$\pm$0.0057& 0.4757\tiny$\pm$0.0115& 0.0571\tiny$\pm$0.0211& 0.5364\tiny$\pm$0.0351& 0.0657\tiny$\pm$0.0197& 0.5452\tiny$\pm$0.0733& 0.0785\tiny$\pm$0.0139& 0.5987\tiny$\pm$0.0448\\
    & vGraph+RS& 0.0559\tiny$\pm$0.0152& 0.5700\tiny$\pm$0.0149& 0.0914\tiny$\pm$0.0089& 0.7046\tiny$\pm$0.0550& 0.1332\tiny$\pm$0.0116& 0.7233\tiny$\pm$0.0461& 0.1573\tiny$\pm$0.0098& 0.7708\tiny$\pm$0.0398\\
    & vGraph+DFS& 0.0601\tiny$\pm$0.0221& 0.5526\tiny$\pm$0.0527& 0.0943\tiny$\pm$0.0104& 0.6614\tiny$\pm$0.0288& 0.1030\tiny$\pm$0.0180& 0.7079\tiny$\pm$0.0584& 0.1272\tiny$\pm$0.0103& 0.7297\tiny$\pm$0.0356\\
    & CLARE+RS& 0.0681\tiny$\pm$0.0110& 0.6623\tiny$\pm$0.2354& 0.0905\tiny$\pm$0.0117& 0.6945\tiny$\pm$0.0277& 0.1171\tiny$\pm$0.0130& 0.7351\tiny$\pm$0.0204& 0.1312\tiny$\pm$0.0113& 0.7621\tiny$\pm$0.0209\\
    & CLARE+DFS& 0.0691\tiny$\pm$0.0131& 0.6843\tiny$\pm$0.0200& 0.0961\tiny$\pm$0.0137& 0.7081\tiny$\pm$0.0162& 0.1008\tiny$\pm$0.0121& 0.7245\tiny$\pm$0.0251& 0.1284\tiny$\pm$0.0122& 0.7548\tiny$\pm$0.0130\\
    & NOCD+RS& 0.0704\tiny$\pm$0.0191& 0.6503\tiny$\pm$0.0960& 0.1022\tiny$\pm$0.0189& 0.6627\tiny$\pm$0.0436& 0.1422\tiny$\pm$0.0111& 0.7445\tiny$\pm$0.0168& 0.1958\tiny$\pm$0.0191& 0.7652\tiny$\pm$0.0399\\
    & NOCD+DFS& 0.0594\tiny$\pm$0.0111& 0.5471\tiny$\pm$0.0283& 0.0965\tiny$\pm$0.0213& 0.6568\tiny$\pm$0.0616& 0.1329\tiny$\pm$0.0218& 0.7352\tiny$\pm$0.0713& 0.1683\tiny$\pm$0.0247& 0.7885\tiny$\pm$0.0542\\
    & DMoN+RS& 0.0681\tiny$\pm$0.0297& 0.6483\tiny$\pm$0.0374& 0.0914\tiny$\pm$0.0209& 0.7197\tiny$\pm$0.0184& 0.1256\tiny$\pm$0.0119& 0.7596\tiny$\pm$0.0325& 0.1449\tiny$\pm$0.0132& 0.7930\tiny$\pm$0.0428\\
    & DMoN+DFS& 0.0702\tiny$\pm$0.0178& 0.6521\tiny$\pm$0.0374& 0.0923\tiny$\pm$0.0184& 0.7258\tiny$\pm$0.0446& 0.1264\tiny$\pm$0.0152& 07481\tiny$\pm$0.0494& 0.1426\tiny$\pm$0.0152& 0.7715\tiny$\pm$0.0358\\
    \cmidrule{2-10}
    & \textsf{META-CODE}\tiny sim& \underline{0.0907}\tiny$\pm$0.0226& \underline{0.7206}\tiny$\pm$0.0473& \underline{0.1572}\tiny$\pm$0.0335& \underline{0.7515}\tiny$\pm$0.0282& \underline{0.2054}\tiny$\pm$0.0188& \underline{0.8287}\tiny$\pm$0.0408& \underline{0.2220}\tiny$\pm$0.0142& \underline{0.8589}\tiny$\pm$0.0428\\
    & \textsf{META-CODE}& {\bf 0.0980}\tiny$\pm$0.0242& {\bf 0.7493}\tiny$\pm$0.0673& {\bf 0.1692}\tiny$\pm$0.0464& {\bf 0.8029}\tiny$\pm$0.0688& {\bf 0.2192}\tiny$\pm$0.0229& {\bf 0.8623}\tiny$\pm$0.0142& {\bf 0.2517}\tiny$\pm$0.0097& {\bf 0.8818}\tiny$\pm$0.0367\\
    \midrule[1pt]
    \multirow{11}*{\rotatebox{90}{Engineering}} &
    BIGCLAM+RS & 0.0000\tiny$\pm$0.0000& 0.1315\tiny$\pm$0.0386 & 0.0000\tiny$\pm$0.0000& 0.2635\tiny$\pm$0.0207 &0.0000\tiny$\pm$0.0000 &0.3015\tiny$\pm$0.0531 &0.0144\tiny$\pm$0.0133 & 0.4659\tiny$\pm$0.0317\\
    & BIGCLAM+DFS& 0.0029\tiny$\pm$0.0066& 0.3460\tiny$\pm$0.0370& 0.0194\tiny$\pm$0.0123& 0.4530\tiny$\pm$0.0674& 0.0406\tiny$\pm$0.0192& 0.5160\tiny$\pm$0.0351& 0.0590\tiny$\pm$0.0088& 0.5980\tiny$\pm$0.0387\\
    & vGraph+RS& 0.0000\tiny$\pm$0.0000& 0.1972\tiny$\pm$0.0282& 0.0000\tiny$\pm$0.0000& 0.3334\tiny$\pm$0.0080& 0.0334\tiny$\pm$0.0042& 0.4702\tiny$\pm$0.0188& 0.0522\tiny$\pm$0.0116& 0.5877\tiny$\pm$0.0913\\
    & vGraph+DFS& 0.0000\tiny$\pm$0.0000& 0.3219\tiny$\pm$0.0084& 0.0281\tiny$\pm$0.0114& 0.4449\tiny$\pm$0.0642& 0.0368\tiny$\pm$0.0054& 0.4811\tiny$\pm$0.0820& 0.0504\tiny$\pm$0.0194& 0.5217\tiny$\pm$0.0137\\
    & CLARE+RS& 0.0109\tiny$\pm$0.0104& 0.2641\tiny$\pm$0.0129& 0.0194\tiny$\pm$0.0117& 0.3747\tiny$\pm$0.0151& 0.0541\tiny$\pm$0.0122& 0.4387\tiny$\pm$0.0226& 0.0674\tiny$\pm$0.0192& 0.4954\tiny$\pm$0.0174\\
    & CLARE+DFS& 0.0114\tiny$\pm$0.0123& 0.2954\tiny$\pm$0.0154& 0.0189\tiny$\pm$0.0110& 0.3528\tiny$\pm$0.0138& 0.0529\tiny$\pm$0.0114& 0.4219\tiny$\pm$0.0245& 0.0546\tiny$\pm$0.0148& 0.5789\tiny$\pm$0.0216\\
    & NOCD+RS& 0.1262\tiny$\pm$0.0219& 0.7081\tiny$\pm$0.0277& 0.2148\tiny$\pm$0.0113& 0.7924\tiny$\pm$0.0367& \underline{0.2875}\tiny$\pm$0.0182& \underline{0.8387}\tiny$\pm$0.0238& \underline{0.3128}\tiny$\pm$0.0175& \underline{0.8529}\tiny$\pm$0.0370\\
    & NOCD+DFS& 0.1466\tiny$\pm$0.0322& 0.7542\tiny$\pm$0.0598& 0.2046\tiny$\pm$0.0185& 0.7760\tiny$\pm$0.0638& 0.2702\tiny$\pm$0.0158& 0.8261\tiny$\pm$0.0304& 0.3038\tiny$\pm$0.0172& 0.8427\tiny$\pm$0.0178\\
    & DMoN+RS& 0.1782\tiny$\pm$0.0276& 0.7562\tiny$\pm$0.0090& 0.2248\tiny$\pm$0.0148& 0.7854\tiny$\pm$0.0432& 0.2534\tiny$\pm$0.0266& 0.8095\tiny$\pm$0.0222& 0.2693\tiny$\pm$0.0265& 0.8242\tiny$\pm$0.0302\\
    & DMoN+DFS& 0.1822\tiny$\pm$0.0195& \underline{0.7691}\tiny$\pm$0.0402& 0.2204\tiny$\pm$0.0062& 0.7898\tiny$\pm$0.0319& 0.2396\tiny$\pm$0.0271& 0.8131\tiny$\pm$0.0432& 0.2431\tiny$\pm$0.0250& 0.8237\tiny$\pm$0.0356\\
    \cmidrule{2-10}
    & \textsf{META-CODE}\tiny sim & \underline{0.1858}\tiny$\pm$0.0258& 0.7584\tiny$\pm$0.0490& \underline{0.2324}\tiny$\pm$0.0246& \underline{0.7941}\tiny$\pm$0.0421& 0.2671\tiny$\pm$0.0466& 0.8105\tiny$\pm$0.0240& 0.3032\tiny$\pm$0.0288&  0.8316\tiny$\pm$0.0456\\
    & \textsf{META-CODE} & {\bf 0.2072}\tiny$\pm$0.0198& {\bf 0.7838}\tiny$\pm$0.0366& {\bf 0.2785}\tiny$\pm$0.0430& {\bf 0.8208}\tiny$\pm$0.0341& {\bf 0.3080}\tiny$\pm$0.0256& {\bf 0.8548}\tiny$\pm$0.0207& {\bf 0.3288}\tiny$\pm$0.0326& {\bf 0.8694}\tiny$\pm$0.0280\\
        \midrule[1pt]
    \multirow{11}*{\rotatebox{90}{Computer Science}}
    & BIGCLAM+RS& 0.0000\tiny$\pm$0.0000& 0.1164\tiny$\pm$0.0104& 0.0000\tiny$\pm$0.0000& 0.2167\tiny$\pm$0.0089& 0.0000\tiny$\pm$0.0000& 0.2914\tiny$\pm$0.0214& 0.0042\tiny$\pm$0.0003& 0.3624\tiny$\pm$0.0312\\
    & BIGCLAM+DFS& 0.0000\tiny$\pm$0.0000 &0.1739\tiny$\pm$0.0213& 0.0000\tiny$\pm$0.0000& 0.2154\tiny$\pm$0.0111& 0.0000\tiny$\pm$0.0000& 0.3468\tiny$\pm$0.0341& 0.0035\tiny$\pm$0.0004& 0.3719\tiny$\pm$0.0272\\
    & vGraph+RS& 0.0017\tiny$\pm$0.0003& 0.3140\tiny$\pm$0.0152& 0.0028\tiny$\pm$0.0024& 0.3379\tiny$\pm$0.0012& 0.0050\tiny$\pm$0.0061& 0.3543\tiny$\pm$0.0323& 0.0106\tiny$\pm$0.0172& 0.3579\tiny$\pm$0.0181\\
    & vGraph+DFS& 0.0042\tiny$\pm$0.0002& 0.2942\tiny$\pm$0.0143& 0.0046\tiny$\pm$0.0004& 0.3151\tiny$\pm$0.0042& 0.0051\tiny$\pm$0.0012& 0.3308\tiny$\pm$0.0073& 0.0056\tiny$\pm$0.0006& 0.3419\tiny$\pm$0.0086\\
    & CLARE+RS& 0.0120\tiny$\pm$0.0118& 0.3754\tiny$\pm$0.0218& 0.0278\tiny$\pm$0.0141& 0.4884\tiny$\pm$0.0284& 0.0528\tiny$\pm$0.0101& 0.5187\tiny$\pm$0.0235& 0.0754\tiny$\pm$0.0147& 0.5478\tiny$\pm$0.0238\\
    & CLARE+DFS& 0.0145\tiny$\pm$0.0111& 0.4085\tiny$\pm$0.0176& 0.0257\tiny$\pm$0.0132& 0.4548\tiny$\pm$0.0213& 0.0492\tiny$\pm$0.0130& 0.4999\tiny$\pm$0.0255& 0.0698\tiny$\pm$0.0127& 0.5265\tiny$\pm$0.0247\\
    & NOCD+RS& 0.2568\tiny$\pm$0.0411& 0.7125\tiny$\pm$0.0226& \underline{0.3724}\tiny$\pm$0.0089& 0.7810\tiny$\pm$0.0144& \underline{0.3972}\tiny$\pm$0.0281& 0.8174\tiny$\pm$0.0173& 0.4205\tiny$\pm$0.0221& 0.8438\tiny$\pm$0.0164\\
    & NOCD+DFS& 0.1741\tiny$\pm$0.0302& 0.6738\tiny$\pm$0.0377& 0.3018\tiny$\pm$0.0118& 0.7473\tiny$\pm$0.0145& 0.3617\tiny$\pm$0.0275& 0.7947\tiny$\pm$0.0173& 0.4074\tiny$\pm$0.0165& 0.8250\tiny$\pm$0.0270\\
    & DMoN+RS& 0.2230\tiny$\pm$0.0656& 0.7569\tiny$\pm$0.0541& 0.2583\tiny$\pm$0.0455& 0.7719\tiny$\pm$0.0533& 0.2914\tiny$\pm$0.0123& 0.8095\tiny$\pm$0.0411& 0.3087\tiny$\pm$0.0379& 0.8128\tiny$\pm$0.0420\\
    & DMoN+DFS& 0.2077\tiny$\pm$0.0550& 0.7570\tiny$\pm$0.0389& 0.2414\tiny$\pm$0.0438& 0.7787\tiny$\pm$0.0375& 0.2784\tiny$\pm$0.0243& 0.7799\tiny$\pm$0.0310& 0.2961\tiny$\pm$0.0210& 0.8018\tiny$\pm$0.0419\\
    \cmidrule{2-10}
    & \textsf{META-CODE}\tiny sim& \underline{0.3138}\tiny$\pm$0.0177& \underline{0.7882}\tiny$\pm$0.0258& 0.3408\tiny$\pm$0.0313& \underline{0.8113}\tiny$\pm$0.0289& 0.3799\tiny$\pm$0.0268& \underline{0.8437}\tiny$\pm$0.0379& \underline{0.4223}\tiny$\pm$0.0261& \underline{0.8595}\tiny$\pm$0.0254\\
    & \textsf{META-CODE}& {\bf 0.3507}\tiny$\pm$0.0204& {\bf 0.8124}\tiny$\pm$0.0381& {\bf 0.4005}\tiny$\pm$0.0169& {\bf 0.8408}\tiny$\pm$0.0322& {\bf 0.4237}\tiny$\pm$0.0360& {\bf 0.8611}\tiny$\pm$0.0480& {\bf 0.4408}\tiny$\pm$0.0223& {\bf 0.8820}\tiny$\pm$0.0389\\
    \midrule[1pt]
    \multirow{12}*{\rotatebox{90}{Medicine}}
    & BIGCLAM+RS & 0.0000\tiny$\pm$0.0000& 0.1656\tiny$\pm$0.0249& 0.0000\tiny$\pm$0.0000& 0.2116\tiny$\pm$0.0318& 0.0000\tiny$\pm$0.0000& 0.2456\tiny$\pm$0.0335& 0.0000\tiny$\pm$0.0000& 0.2908\tiny$\pm$0.0321\\
    & BIGCLAM+DFS & 0.0000\tiny$\pm$0.0000& 0.1823\tiny$\pm$0.0303& 0.0000\tiny$\pm$0.0000& 0.2438\tiny$\pm$0.0335& 0.0000\tiny$\pm$0.0000& 0.2845\tiny$\pm$0.0297& 0.0000\tiny$\pm$0.0000& 0.3178\tiny$\pm$0.0340\\
    & vGraph+RS & 0.0000\tiny$\pm$0.0000& 0.2715\tiny$\pm$0.0265& 0.0000\tiny$\pm$0.0000& 0.2987\tiny$\pm$0.0292& 0.0000\tiny$\pm$0.0000& 0.3247\tiny$\pm$0.0374& 0.0000\tiny$\pm$0.0000& 0.3426\tiny$\pm$0.0302\\
    & vGraph+DFS & 0.0000\tiny$\pm$0.0000& 0.2778\tiny$\pm$0.0224& 0.0000\tiny$\pm$0.0000& 0.2886\tiny$\pm$0.0315& 0.0000\tiny$\pm$0.0000& 0.3138\tiny$\pm$0.0301& 0.0000\tiny$\pm$0.0000& 0.3315\tiny$\pm$0.0311\\
    & CLARE+RS & 0.0145\tiny$\pm$0.0117& 0.2616\tiny$\pm$0.0223& 0.0303\tiny$\pm$0.0168& 0.3728\tiny$\pm$0.0209& 0.0584\tiny$\pm$0.0147& 0.4795\tiny$\pm$0.0197& 0.0754\tiny$\pm$0.0213& 0.5121\tiny$\pm$0.0374\\
    & CLARE+DFS & 0.0207\tiny$\pm$0.0115& 0.3077\tiny$\pm$0.0306& 0.0375\tiny$\pm$0.0146& 0.3884\tiny$\pm$0.0239& 0.0529\tiny$\pm$0.0129& 0.4788\tiny$\pm$0.0251& 0.0645\tiny$\pm$0.0194& 0.5006\tiny$\pm$0.0306\\
    & NOCD+RS &0.2431\tiny$\pm$0.0179& 0.7090\tiny$\pm$0.0353& \underline{0.3295}\tiny$\pm$0.0364& 0.7789\tiny$\pm$0.0197& 0.3375\tiny$\pm$0.0317& 0.8011\tiny$\pm$0.0461& 0.3478\tiny$\pm$0.0148& 0.8134\tiny$\pm$0.0225\\
    & NOCD+DFS& 0.2458\tiny$\pm$0.0509& 0.6899\tiny$\pm$0.0408& 0.2833\tiny$\pm$0.0133& 0.7509\tiny$\pm$0.0212& 0.3288\tiny$\pm$0.0251& 0.7946\tiny$\pm$0.0109& 0.3416\tiny$\pm$0.0129& 0.8013\tiny$\pm$0.0190\\
    & DMoN+RS& 0.1172\tiny$\pm$0.0201& 0.6725\tiny$\pm$0.0226& 0.1215\tiny$\pm$0.0183& 0.7130\tiny$\pm$0.0183& 0.1447\tiny$\pm$0.0152& 0.7384\tiny$\pm$0.0196& 0.1545\tiny$\pm$0.0239& 0.7449\tiny$\pm$0.0257\\
    &DMoN+DFS&  0.1089\tiny$\pm$0.0198& 0.6811\tiny$\pm$0.0359& 0.1466\tiny$\pm$0.0258& 0.7382\tiny$\pm$0.0294& 0.1559\tiny$\pm$0.0215& 0.7466\tiny$\pm$0.0374& 0.1790\tiny$\pm$0.0232& 0.7568\tiny$\pm$0.0373\\
    \cmidrule{2-10}
    & \textsf{META-CODE}\tiny sim& \underline{0.2568}\tiny$\pm$0.0204& \underline{0.7521}\tiny$\pm$0.0346& 0.3245\tiny$\pm$0.0224& \underline{0.7963}\tiny$\pm$0.0369& \underline{0.3474}\tiny$\pm$0.0240& \underline{0.8125}\tiny$\pm$0.0384& \underline{0.3584}\tiny$\pm$0.0216& \underline{0.8264}\tiny$\pm$0.0348\\
    & \textsf{META-CODE}& {\bf 0.2733}\tiny$\pm$0.0357& {\bf 0.7931}\tiny$\pm$0.0363& {\bf 0.3527}\tiny$\pm$0.0289& {\bf 0.8384}\tiny$\pm$0.0311& {\bf 0.3583}\tiny$\pm$0.0271& {\bf 0.8412}\tiny$\pm$0.0394& {\bf 0.3653}\tiny$\pm$0.0248& {\bf 0.8497}\tiny$\pm$0.0307\\
    \bottomrule[1pt]
  \end{tabular}
  \vspace{-2.5em}
\end{table*}

In this subsection, we describe the comprehensive experiments conducted to answer RQ1--RQ9. To address RQ1, we provide the experiment results on all datasets. To address RQ2--RQ4, RQ6, and RQ7, we detail the results on Facebook and Engineering since the results on the other datasets have similar tendencies.

\subsubsection{Comparison With Benchmark Methods ({\bf RQ1})}
\label{sec:RQ1}
A performance comparison of our \textsf{META-CODE} methods as well as five competitive overlapping community detection methods, each combined with two sampling strategies, is presented in Table \ref{tab:comparison}. The comparison is made with respect to two performance metrics, NMI and Avg$\text{F}_1$, across five real-world datasets where a different percentage of nodes are queried among $N$ nodes. We note that the hyperparameters in all the aforementioned methods are tuned differently according to each individual dataset to obtain the best performance. We make the following insightful observations:

\begin{itemize}
    \item Our \textsf{META-CODE} method consistently and significantly outperforms all the benchmark methods regardless of the datasets and the performance metrics.

    \item The second-best performer tends to be \textsf{META-CODE}{\tiny sim}. \textsf{META-CODE}{\tiny sim} is superior to the five competitive benchmark methods for almost all cases since it contains modules with sophisticated design, which include 1) network inference with the aid of EC-SiamNet and 2) GNN-aided community-affiliation embedding using our reconstruction loss. However, \textsf{META-CODE}{\tiny sim} is always inferior to the original \textsf{META-CODE} due to the lack of network exploration via judicious node queries and iterative steps. In other words, \textsf{META-CODE}{\tiny sim} overlooks not only the potential of query node selection to expedite network exploration but also the gradual increase of performance in community detection through the iterative process.  

    \item Among the competitive benchmark methods, the best competitor tends to be NOCD with RS in most cases since NOCD also generates community-affiliation embeddings via GNNs with the aid of node metadata, which is capable of precisely discovering communities. In comparison with NOCD, the proposed \textsf{META-CODE} and \textsf{META-CODE}{\tiny sim} methods use reconstruction loss, which takes into account both structure--community and metadata--community relationships to generate community-affiliation embeddings, thereby resulting in better performance than NOCD. 
    
    \item DMoN leverages modularity for training GNNs, where modularity relies heavily on the network structure. This may result in comparatively lower performance than that of NOCD, especially when only a few nodes are queried.

    \item The performance gap between \textsf{META-CODE} ($X$) and \textsf{META-CODE}{\tiny sim} ($Y$) is the largest when the Engineering dataset is used; the maximum improvement rate of 19.84\% is achieved in terms of NMI, where the improvement rate (\%) is given by $\frac{{X - Y}}{Y} \times 100 $. Additionally, the maximum improvement rate between our \textsf{META-CODE} method and the best competitive benchmark method, NOCD, is 65.55\% in terms of NMI when the Facebook dataset is used.

    \item The improvement rates of \textsf{META-CODE} in terms of NMI and Avg$\text{F}_1$ tend to decrease as the percentage of queried nodes increases. For instance, in the Computer Science dataset, the NMI improvement rates are 14.20\%, 5.79\%, and 4.03\% when the percentage of queried nodes increases from 10\% to 20\%, 20\% to 30\%, and 30\% to 40\%, respectively. This is due to the fact that, as the percentage of queried nodes increases, the size of the unexplored part in the network diminishes and the explored network converges to the underlying true network.

    \item Let us compare two sampling strategies for network exploration used in the benchmark methods. When a small number of nodes are queried (i.e., the explored network is small), combining each benchmark method with DFS tends to result in better performance compared to the case using RS. However, when a sufficient number of nodes are queried, using RS for network exploration tends to result in better performance. For example, on Facebook, vGraph with DFS is superior to vGraph with RS when 10--20\% of nodes are queried, while vGraph with RS performs better when more than 30\% of nodes are queried. This is because DFS can {\it locally} explore the subnetwork in a single community faster, which is beneficial to the early stage of community detection. On the other hand, as the number of queried nodes increases, the local network in a community may be almost explored and it is better to explore other parts of the underlying network. This supports the claim that the node query strategy in \textsf{META-CODE} is designed in such a way that nodes distributed over {\it diverse} communities are selected for faster exploration.

\end{itemize}

\begin{figure}[t!]
\pgfplotsset{footnotesize,samples=10}
\centering
\begin{tikzpicture}
\begin{axis}[
legend columns=6,
legend entries={\textsf{META-CODE},RS,DFS},
legend to name=named,
xlabel= (a) Facebook, ylabel = $N_{ex}$,  width = 3.7cm, height = 3.7cm,
xmin=0.1,xmax=0.4,ymin=150,ymax=230,
xticklabel={\pgfmathparse{\tick*100}\pgfmathprintnumber{\pgfmathresult}\%},
ytick={150,175,200,225},
]
    \addplot+[color=black] coordinates{(0.1, 191)(0.2, 215)(0.3, 223)(0.4, 225)};
    \addplot+[color=orange] coordinates{(0.1, 177)(0.2, 201)(0.3, 212)(0.4, 218)};
    \addplot+[color=purple] coordinates{(0.1, 156)(0.2, 189)(0.3, 196)(0.4, 199)};
\end{axis}
\end{tikzpicture}
\begin{tikzpicture}
\begin{axis}[
xlabel=(b) Engineering, ylabel = $N_{ex}$,  width = 3.7cm, height = 3.7cm,
xmin=0.1,xmax=0.4,ymin=5000,ymax=15000,
xticklabel={\pgfmathparse{\tick*100}\pgfmathprintnumber{\pgfmathresult}\%},
ytick={5000,8000,11000,14000}]
    \addplot+[color=black] coordinates{(0.1, 7153)(0.2, 10470)(0.3, 12872)(0.4, 14285)};
    \addplot+[color=orange] coordinates{(0.1, 6656)(0.2, 10312)(0.3, 12340)(0.4, 13465)};
    \addplot+[color=purple] coordinates{(0.1, 5766)(0.2, 8311)(0.3, 10085)(0.4, 11436)};
\end{axis}
\end{tikzpicture}
\\
\ref{named}
\captionsetup{skip=0pt}
\caption{The number of explored nodes, $N_\text{ex}$, according to different percentages (\%) of queried nodes among $N$ nodes on Facebook and Engineering.}
\label{fig:queried_nodes}
\vspace{-2.0em}
\end{figure}

\subsubsection{Growth of Explored Subnetworks ({\bf RQ2})} 
\label{sec:RQ3}
To empirically validate the effectiveness of our node query strategy, we conduct experiments on Facebook and Engineering by plotting the number of explored nodes, denoted as $N_\text{ex}$, versus the percentages of queried nodes in Fig. \ref{fig:queried_nodes}. We compare our query strategy with two cases in which two sampling strategies (i.e., RS and DFS) are employed for network exploration in \textsf{META-CODE} rather than the strategy proposed in (\ref{eq:strategy}). Our findings are as follows: 

\begin{table}[t]\centering
\setlength\tabcolsep{4.0pt}
\scriptsize
\captionsetup{skip=0pt}
  \caption{AUC scores for the inferred network $\mathcal{G}^{(t)}$ from \textsf{META-CODE-4} and \textsf{META-CODE} vs. the underlying network $\mathcal{G}$.}
  \label{tab:AUC}
  \begin{tabular}{c|c|cccl}
    \toprule[1pt]
           Dataset & Queried percentage &\textsf{META-CODE-4} & \textsf{META-CODE} & Gain& \\
    \midrule[1pt]
    \multirow{4}*{{Facebook}} 
    & 10\% & 0.6015& {\bf 0.6823}& 13.43\% \\
    & 20\% & 0.6681& {\bf 0.7304}& 9.32\% \\
    & 30\% & 0.7310& {\bf 0.7864}& 7.57\% \\
    & 40\% & 0.7799& {\bf 0.8302} & 6.44\%\\
    \midrule[1pt]
    \multirow{4}*{{Engineering}} 
    & 10\% & 0.5744& {\bf 0.6220}& 8.28\% \\
    & 20\% & 0.6408& {\bf 0.6794}& 6.02\% \\
    & 30\% & 0.6907& {\bf 0.7308}& 4.89\% \\
    & 40\% & 0.7495& {\bf 0.7771} & 3.68\%\\

    \bottomrule[1pt]
  \end{tabular}
\vspace{-2.0em}
\end{table}

\begin{itemize}
    \item Our query strategy based on the discovered community affiliations is consistently more beneficial compared to other samplings as it leads to the faster exploration of the unknown parts of the underlying network.

    \item The growth rate of $N_\text{ex}$ for RS is even higher than that for DFS since querying nodes away from explored subnetworks via RS (instead of expanding the explored subnetwork via DFS) is more effective as the number of queried nodes increases.
\end{itemize}

\subsubsection{Convergence of the Inferred Network ({\bf RQ3})} 
\label{sec:RQ4}

It is worth analyzing the convergence of the inferred network to the underlying true network, which is crucial for finding more precise community-affiliation embeddings through iterations. In Table \ref{tab:AUC}, we present the area under the curve (AUC) score to measure the similarity between the inferred network $\mathcal{G}^{(t)}$ in both \textsf{META-CODE-4} (mentioned in Section \ref{sec:RQ2})) and \textsf{META-CODE} and the underlying network $\mathcal{G}$ for different percentages of queried nodes when the Facebook and Engineering datasets are used. The AUC score between $\mathcal{G}^{(t)}$ and $\mathcal{G}$ can be computed by comparing the two corresponding adjacency matrices, where the adjacency matrices from $\mathcal{G}^{(t)}$ and $\mathcal{G}$ can be seen as the predicted outcome and the true label, respectively. A higher AUC indicates a greater similarity between the two networks being compared. Our findings are as follows:

\begin{itemize}
    \item The performance gap between \textsf{META-CODE-4} and \textsf{META-CODE} is largest when the Facebook dataset is used; the maximum improvement rate of 13.43\% is achieved in terms of AUC score when 10\% of nodes are queried.

    \item The AUC score between $\mathcal{G}^{\left( t \right)} $ and $\mathcal{G}$ tends to increase as the percentage of node queries grows. This suggests that EC-SiamNet is effective in accurately inferring edges, leading to a fast convergence of the inferred network to the underlying true network, as noted in Remark \ref{remark:con}.
\end{itemize}

\begin{table*}[th]\centering
\setlength\tabcolsep{5.0pt}
\scriptsize
\captionsetup{skip=0pt}
  \caption{Performance comparison of \textsf{META-CODE} using three different GNN backbones. The best and second-best performers are bolded and underlined, respectively.}
  \label{tab:gnns}
  \resizebox{\textwidth}{!}{
  \begin{tabular}{cc|cc|cc|cc|ccl}
    \toprule[1pt]
    \multicolumn{2}{c|}{}&\multicolumn{2}{|c|}{10\% queried nodes}&\multicolumn{2}{c|}{20\% queried nodes}&\multicolumn{2}{c|}{30\% queried nodes}&\multicolumn{2}{c}{40\% queried nodes}\\
    \cmidrule{1-10}
           Dataset& Method & NMI& Avg$\text{F}_1$ & NMI& Avg$\text{F}_1$ & NMI& Avg$\text{F}_1$ & NMI& Avg$\text{F}_1$ \\
    \midrule[1pt]
    \multirow{3}*{Facebook} 
    & \textsf{META-CODE (GraphSAGE)}& 0.0951\tiny$\pm$0.0123& 0.7312\tiny$\pm$0.0309& 0.1629\tiny$\pm$0.0265& 0.7962\tiny$\pm$0.0212& 0.2125\tiny$\pm$0.0198& 0.8518\tiny$\pm$0.0269& 0.2418\tiny$\pm$0.0298& 0.8745\tiny$\pm$0.0213\\
    & \textsf{META-CODE (GAT)}& \underline{0.0964}\tiny$\pm$0.0127& \underline{0.7347}\tiny$\pm$0.0317& \underline{0.1644}\tiny$\pm$0.0277& \underline{0.7997}\tiny$\pm$0.0215& \underline{0.2142}\tiny$\pm$0.0201& \underline{0.8543}\tiny$\pm$0.0272& 0.2441\tiny$\pm$0.0304& \underline{0.8770}\tiny$\pm$0.0216\\
    &{\textsf{META-CODE (GCN)}}& {\bf 0.0980}\tiny$\pm$0.0242& {\bf 0.7493}\tiny$\pm$0.0673& {\bf 0.1692}\tiny$\pm$0.0464& {\bf 0.8029}\tiny$\pm$0.0688& {\bf 0.2192}\tiny$\pm$0.0229& {\bf 0.8623}\tiny$\pm$0.0142& {\bf 0.2517}\tiny$\pm$0.0097& {\bf 0.8818}\tiny$\pm$0.0367\\
    \midrule[1pt]
    \multirow{3}*{Engineering} 
    & \textsf{META-CODE (GraphSAGE)}& 0.1924\tiny$\pm$0.0301 & 0.7418\tiny$\pm$0.0310 & 0.2610\tiny$\pm$0.0238 & 0.8123\tiny$\pm$0.0195 & 0.2990\tiny$\pm$0.0187 & 0.8451\tiny$\pm$0.0365 & 0.3187\tiny$\pm$0.0235 & 0.8602\tiny$\pm$0.0302 \\
    & \textsf{META-CODE (GAT)}& \underline{0.1968}\tiny$\pm$0.0315& \underline{0.7485}\tiny$\pm$0.0321& \underline{0.2649}\tiny$\pm$0.0245& \underline{0.8159}\tiny$\pm$0.0207& \underline{0.3022}\tiny$\pm$0.0199& \underline{0.8502}\tiny$\pm$0.0379& \underline{0.3214}\tiny$\pm$0.0244& \underline{0.8633}\tiny$\pm$0.0316\\
    &\textsf{META-CODE (GCN)}&  {\bf 0.2072}\tiny$\pm$0.0198& {\bf 0.7838}\tiny$\pm$0.0866& {\bf 0.2785}\tiny$\pm$0.0430& {\bf 0.8208}\tiny$\pm$0.0341& {\bf 0.3080}\tiny$\pm$0.0256& {\bf 0.8548}\tiny$\pm$0.0207& {\bf 0.3288}\tiny$\pm$0.0326& {\bf 0.8692}\tiny$\pm$0.0580\\
    \bottomrule[1pt]
  \end{tabular}
  }
\vspace{-2.0em}
\end{table*}

\subsubsection{Comparison among Different GNN Backbones ({\bf RQ4})}

When three different GNN models, including GCN \cite{kipf2016semi}, GAT \cite{velivckovic2017graph}, and GraphSAGE \cite{hamilton2017inductive}, are employed as a GNN backbone of our \textsf{META-CODE} to generate community-affiliation embeddings, a performance comparison is presented in Table \ref{tab:gnns}. The comparison is made in terms of two performance metrics, NMI and Avg$\text{F}_1$, across the Facebook and Engineering datasets, where varying percentages of nodes are queried among $N$ nodes. The results demonstrate that, although the use of GCN tends to consistently outperform the other two, the performance gap among GNN encoders is indeed negligible. This implies that, as long as the message passing mechanism, aggregating information from neighboring nodes, is concerned, any GNN backbone can be applied to our \textsf{META-CODE} method while guaranteeing satisfactory performance.

\subsubsection{Comparison in terms of Modularity ({\bf RQ5})}

Modularity \cite{newman2006modularity, nicosia2009extending} is a metric that evaluates the quality of a network's partition into communities by comparing the observed density of edges within communities to the density expected in a random graph with the same degree distribution. The modularity $Q$ is defined as:
\[
Q = \frac{1}{{2\left| \mathcal{E} \right|}}\sum\limits_{u,v} {\left[ {A_{u,v}  - \frac{{k_u k_v }}{{\left| \mathcal{E} \right|}}} \right]\delta \left( {c_u ,c_v } \right)},
\]
where $A_{u,v}$ indicates the connection between two nodes $u$ and $v$ (i.e., $A_{u,v}=1$ if node $u$ and $v$ are connected, otherwise $A_{u,v}=0$); $k_u$ and $k_v$ are the degrees of nodes $u$ and $v$; $c_u$ and $c_v$ denote the communities of nodes $u$ and $v$; and $\delta \left( {c_u ,c_v } \right)$ is 1 if nodes $u$ and $v$ belong to the same community, and 0 otherwise. Modularity quantifies how well the community structure captures the network's internal organization by focusing on the intra-community edge density relative to the inter-community edge density.

Given that modularity is highly influenced by the network structure and is sensitive to substantial variations in its structural composition, we evaluate the performance by fixing the percentage of queried nodes to 40\%. A performance comparison of our \textsf{META-CODE} method and five state-of-the-art overlapping community detection methods is presented in Table \ref{tab:modularity} in terms of modularity, without relying on real community labels for evaluation, using the Engineering dataset. The results demonstrate that \textsf{META-CODE} consistently achieves higher modularity scores compared to other benchmark methods, indicating its effectiveness in discovering meaningful community structures even without ground truth communities.

\begin{table}[t]\centering
\setlength\tabcolsep{5.0pt}
\scriptsize
\captionsetup{skip=0pt}
  \caption{The community detection performance in terms of Modularity on the Engineering dataset. The best and second-best performers are bolded and underlined, respectively.}
  \label{tab:modularity}
  \begin{tabular}{cc|c}
    \toprule[1pt]
    Dataset&Method&40\% queried nodes\\
    \midrule[1pt]
    \multirow{12}*{\rotatebox{90}{Engineering}} 
    & BIGCLAM+RS &0.1215\tiny$\pm$0.0275\\
    & BIGCLAM+DFS & 0.1107\tiny$\pm$0.0283\\
    & vGraph+RS & 0.2224\tiny$\pm$0.0204\\
    & vGraph+DFS & 0.2174\tiny$\pm$0.0213\\
    & CLARE+RS & 0.2318\tiny$\pm$0.0166\\
    & CLARE+DFS & 0.2212\tiny$\pm$0.0124\\
    & NOCD+RS & 0.3768\tiny$\pm$0.0103\\
    & NOCD+DFS & 0.3672\tiny$\pm$0.0098\\
    & DMoN+RS & 0.3863\tiny$\pm$0.0187\\
    & DMoN+DFS & 0.3744\tiny$\pm$0.0144\\
    \cmidrule{2-3}
    & \textsf{META-CODE}\tiny sim& \underline{0.4121}\tiny$\pm$0.0162\\
    &{\textsf{META-CODE}}& {\bf 0.4332}\tiny$\pm$0.0251\\
    \bottomrule[1pt]
  \end{tabular}
\vspace{-2.0em}
\end{table}

\subsubsection{Impact of Components in \textsf{META-CODE} ({\bf RQ6})}
\label{sec:RQ2}
To discover what role each component plays in the success of the proposed \textsf{META-CODE} method, we conduct an ablation study by removing or replacing each module in our method.

\begin{itemize}
    \item \textsf{META-CODE}: This corresponds to the original \textsf{META-CODE} method.

    \item \textsf{META-CODE-1}: The GNN model in \textsf{META-CODE} is replaced with BIGCLAM \cite{yang2013overlapping} to generate community-affiliation embeddings.

    \item \textsf{META-CODE-2}: The GNN model in \textsf{META-CODE} is replaced with MLP to generate community-affiliation embeddings.

    \item \textsf{META-CODE-3}: The node query strategy in \textsf{META-CODE} is replaced with RS for network exploration.

    \item \textsf{META-CODE-4}: The network inference step in \textsf{META-CODE} is removed, which corresponds to our prior work in the short paper \cite{hou2022meta}.
\end{itemize}

A performance comparison of the original \textsf{META-CODE} and its four variants is presented in Table \ref{tab:ablation} with respect to NMI and Avg$\text{F}_1$ using the Facebook and Engineering datasets. Our findings are as follows:

\begin{itemize}
    \item The original \textsf{META-CODE} method always exhibits potential gains over the other variants, which demonstrates that each module plays a critical role together in discovering communities.
    \item The performance gap between \textsf{META-CODE} and \textsf{META-CODE-2} is much higher compared to the other variants for both datasets.
    
    \item \textsf{META-CODE-1} outperforms \textsf{META-CODE-2}, which can be attributed to the fact that BIGCLAM was originally designed for community detection, whereas MLP fails to precisely capture the structure--community relationship.

    \item \textsf{META-CODE} indeed achieves remarkable gains over \textsf{META-CODE-4} (i.e., our short paper \cite{hou2022meta}), which implies that the network inference step is essential in achieving more accurate community detection results. Compared to \textsf{META-CODE-4}, \textsf{META-CODE} leverages the structural information acquired from network exploration to the maximum, resulting in superior community detection performance.
\end{itemize}

\begin{table*}[t]\centering
\setlength\tabcolsep{5.0pt}
\scriptsize
\captionsetup{skip=0pt}
  \caption{Ablation study on Facebook and Engineering. The best and second-best performers are bolded and underlined, respectively.}
  \label{tab:ablation}
  \begin{tabular}{cc|cc|cc|cc|ccl}
    \toprule[1pt]
    \multicolumn{2}{c|}{}&\multicolumn{2}{|c|}{10\% queried nodes}&\multicolumn{2}{c|}{20\% queried nodes}&\multicolumn{2}{c|}{30\% queried nodes}&\multicolumn{2}{c}{40\% queried nodes}\\
    \cmidrule{1-10}
           Dataset& Method & NMI& Avg$\text{F}_1$ & NMI& Avg$\text{F}_1$ & NMI& Avg$\text{F}_1$ & NMI& Avg$\text{F}_1$ \\
    \midrule[1pt]
    \multirow{5}*{\rotatebox{90}{Facebook}} 
    & \textsf{META-CODE-1} &0.0783\tiny$\pm$0.0199& 0.5562\tiny$\pm$0.0151& 0.1151\tiny$\pm$0.0312& 0.6230\tiny$\pm$0.0137& 0.1442\tiny$\pm$0.0329& 0.6711\tiny$\pm$0.0361& 0.1647\tiny$\pm$0.0081& 0.8033\tiny$\pm$0.0446\\
    & \textsf{META-CODE-2}& 0.0669\tiny$\pm$0.0187& 0.5335\tiny$\pm$0.0170& 0.0755\tiny$\pm$0.0352& 0.5846\tiny$\pm$0.0247& 0.0981\tiny$\pm$0.0255& 0.6352\tiny$\pm$0.0394& 0.1368\tiny$\pm$0.0396& 0.6958\tiny$\pm$0.0641\\
    & \textsf{META-CODE-3}& \underline{0.0814}\tiny$\pm$0.0321& \underline{0.7203}\tiny$\pm$0.0516& \underline{0.1588}\tiny$\pm$0.0245& \underline{0.7941}\tiny$\pm$0.0280& \underline{0.2108}\tiny$\pm$0.0197& \underline{0.8098}\tiny$\pm$0.0696& \underline{0.2487}\tiny$\pm$0.0211& \underline{0.8709}\tiny$\pm$0.0231\\
    & \textsf{META-CODE-4}& 0.0791\tiny$\pm$0.0198& 0.6842\tiny$\pm$0.0343& 0.1132\tiny$\pm$0.0167& 0.7221\tiny$\pm$0.0567& 0.1578\tiny$\pm$0.0169& 0.7923\tiny$\pm$0.0493& 0.2003\tiny$\pm$0.0164& 0.8218\tiny$\pm$0.0085\\
    &{\textsf{META-CODE}}& {\bf 0.0980}\tiny$\pm$0.0242& {\bf 0.7493}\tiny$\pm$0.0673& {\bf 0.1692}\tiny$\pm$0.0464& {\bf 0.8029}\tiny$\pm$0.0688& {\bf 0.2192}\tiny$\pm$0.0229& {\bf 0.8623}\tiny$\pm$0.0142& {\bf 0.2517}\tiny$\pm$0.0097& {\bf 0.8818}\tiny$\pm$0.0367\\
    \midrule[1pt]
    \multirow{5}*{\rotatebox{90}{Engineering}} 
    & \textsf{META-CODE-1} & 0.1371\tiny$\pm$0.0108& 0.5729\tiny$\pm$0.0364& 0.1532\tiny$\pm$0.0276& 0.6031\tiny$\pm$0.0219& 0.1887\tiny$\pm$0.0281& 0.6598\tiny$\pm$0.0210& 0.2145\tiny$\pm$0.0294& 0.6946\tiny$\pm$0.0394\\
    & \textsf{META-CODE-2} &0.0921\tiny$\pm$0.0135& 0.5245\tiny$\pm$0.0271& 0.1265\tiny$\pm$0.0217& 0.5725\tiny$\pm$0.0428& 0.1629\tiny$\pm$0.0297& 0.6214\tiny$\pm$0.0239& 0.1968\tiny$\pm$0.0204& 0.6611\tiny$\pm$0.0255\\
    & \textsf{META-CODE-3}& \underline{0.1874}\tiny$\pm$0.0237& \underline{0.7619}\tiny$\pm$0.0331& \underline{0.2584}\tiny$\pm$0.0101& \underline{0.8104}\tiny$\pm$0.0238& \underline{0.2934}\tiny$\pm$0.0274& \underline{0.8487}\tiny$\pm$0.0134& \underline{0.3207}\tiny$\pm$0.0190& \underline{0.8602}\tiny$\pm$0.0207\\
    & \textsf{META-CODE-4}& 0.1575\tiny$\pm$0.0371& 0.7613\tiny$\pm$0.0193& 0.2177\tiny$\pm$0.0207& 0.8079\tiny$\pm$0.0193& 0.2921\tiny$\pm$0.0169& 0.8431\tiny$\pm$0.0177& 0.3191\tiny$\pm$0.0302& 0.8581\tiny$\pm$0.0175\\
    &{\textsf{META-CODE}}&  {\bf 0.2072}\tiny$\pm$0.0198& {\bf 0.7838}\tiny$\pm$0.0866& {\bf 0.2785}\tiny$\pm$0.0430& {\bf 0.8208}\tiny$\pm$0.0341& {\bf 0.3080}\tiny$\pm$0.0256& {\bf 0.8548}\tiny$\pm$0.0207& {\bf 0.3288}\tiny$\pm$0.0326& {\bf 0.8692}\tiny$\pm$0.0580\\
    \bottomrule[1pt]
  \end{tabular}
\vspace{-2.0em}
\end{table*}

\subsubsection{Effect of Hyperparameters ({\bf RQ7})}

Figs. \ref{fig:RQ5_eta} and \ref{fig:RQ5_lambda} illustrate the impact of two key hyperparameters, $\eta$ and $\lambda$, on the performance of \textsf{META-CODE} in terms of the NMI when the Facebook and Engineering datasets are used. When a hyperparameter varies so that its effect is clearly revealed, another parameter is set to the pivot value addressed in Section V-D. Our findings are as follows.

\begin{itemize}
    \item {\bf The effect of} $\eta $: From Fig \ref{fig:RQ5_eta}, the maximum NMI is achieved at $\eta = 1.5$ and $\eta = 1.0$ on Facebook and Engineering, respectively. When $\eta  = 0 $, our loss function in (\ref{eq:loss}) boils down to the case in \cite{shchur2019overlapping}, which only takes into account the structure--community relationship without node metadata, leading to unsatisfactory performance. On the other hand, setting $\eta$ to a high value leads to relatively low performance for both datasets since the metadata--community relationship is over-emphasized during training. Hence, it is crucial to suitably determine the value of $\eta$ in guaranteeing satisfactory performance.

    \item {\bf The effect of} $\lambda $: From Fig \ref{fig:RQ5_lambda}, the maximum NMI is achieved at $\lambda = 2.0$ and $\lambda = 3.0$ on Facebook and Engineering, respectively. When $\lambda = 0 $, our node query strategy in (\ref{eq:strategy}) focuses only on selecting nodes belonging to more communities. In contrast, setting $\lambda$ to a high value leads to a negative effect on performance because it over-emphasizes the selection of nodes distributed over diverse communities. Thus, it is important to appropriately determine the value of $\lambda$ depending on the dataset.
    
\end{itemize}

\begin{figure}[t!]
\pgfplotsset{footnotesize,samples=10}
\centering
\begin{tikzpicture}
\begin{axis}[
legend columns=6,
legend entries={10\%,20\%,30\%,40\%},
legend to name=named,
xlabel= (a) Facebook, ylabel = NMI,  width = 3.7cm, height = 3.7cm,
xmin=0,xmax=2.5,ymin=0.08,ymax=0.29,
xtick={0,0.5,1.0,1.5,2.0,2.5},
ytick={0.1,0.14,0.20,0.27}]
    \addplot+[color=fig_6_color_1] coordinates{(0, 0.2584)(0.5, 0.2606)(1.0, 0.2647)(1.5, 0.2680)(2.0, 0.2676)(2.5, 0.2546)};
    \addplot+[color=fig_6_color_2] coordinates{(0, 0.2025)(0.5, 0.2091)(1.0, 0.2185)(1.5, 0.2234)(2.0, 0.2217)(2.5, 0.2106)};
    \addplot+[color=fig_6_color_3] coordinates{(0, 0.1665)(0.5, 0.1700)(1.0, 0.1737)(1.5, 0.1778)(2.0, 0.1753)(2.5, 0.1690)};
    \addplot+[color=fig_6_color_4] coordinates{(0, 0.0913)(0.5, 0.0963)(1.0, 0.1003)(1.5, 0.1022)(2.0, 0.0992)(2.5, 0.0981)};
\end{axis}
\end{tikzpicture}
\begin{tikzpicture}
\begin{axis}[
xlabel=(b) Engneering, ylabel = NMI,  width = 3.7cm, height = 3.7cm,
xmin=0,xmax=2.5,ymin=0.18,ymax=0.34,
xtick={0,0.5,1.0,1.5,2.0,2.5},
ytick={0.18,0.23,0.28,0.34}]
    \addplot+[color=fig_6_color_1] coordinates{(0, 0.3091)(0.5, 0.3168)(1.0, 0.3224)(1.5, 0.3186)(2.0, 0.3177)(2.5, 0.3104)};
    \addplot+[color=fig_6_color_2] coordinates{(0, 0.2859)(0.5, 0.3036)(1.0, 0.3091)(1.5, 0.3066)(2.0, 0.3047)(2.5, 0.3015)};
    \addplot+[color=fig_6_color_3] coordinates{(0, 0.26099)(0.5, 0.26316)(1.0, 0.26849)(1.5, 0.26720)(2.0, 0.26348)(2.5, 0.26289)};
    \addplot+[color=fig_6_color_4] coordinates{(0, 0.1925)(0.5, 0.1979)(1.0, 0.2064)(1.5, 0.2041)(2.0, 0.2010)(2.5, 0.1993)};

\end{axis}
\end{tikzpicture}
\\
\ref{named}
\captionsetup{skip=1pt}
\caption{NMI according to different $\eta$'s in \textsf{META-CODE} on Facebook and Engineering when different percentages of nodes are queried.}
\label{fig:RQ5_eta}
\vspace{-1.0em}
\end{figure}

\begin{figure}[t!]
\pgfplotsset{footnotesize,samples=10}
\centering
\begin{tikzpicture}
\begin{axis}[
legend columns=6,
legend entries={10\%,20\%,30\%,40\%},
legend to name=named,
xlabel= (a) Facebook, ylabel = NMI,  width = 3.7cm, height = 3.7cm,
xmin=0,xmax=4,ymin=0.08,ymax=0.27,
xtick={0,1,2,3,4},
ytick={0.1,0.14,0.20,0.27}]
    \addplot+[color=fig_7_color_1] coordinates{(0, 0.24114)(1, 0.24816)(2, 0.25337)(3, 0.24562)(4, 0.23032)};
    \addplot+[color=fig_7_color_2] coordinates{(0, 0.1925)(1, 0.20773)(2, 0.21507)(3, 0.2084)(4, 0.19456)};
    \addplot+[color=fig_7_color_3] coordinates{(0, 0.1600)(1, 0.16288)(2, 0.16920)(3, 0.16328)(4, 0.15811)};
    \addplot+[color=fig_7_color_4] coordinates{(0, 0.0913)(1, 0.09704)(2, 0.09801)(3, 0.09760)(4, 0.08909)};
\end{axis}
\end{tikzpicture}
\begin{tikzpicture}
\begin{axis}[
xlabel=(b) Engineering, ylabel = NMI,  width = 3.7cm, height = 3.7cm,
xmin=0,xmax=4,ymin=0.18,ymax=0.34,
xtick={0,1,2,3,4},
ytick={0.18,0.23,0.28,0.34}]
    \addplot+[color=fig_7_color_1] coordinates{(0, 0.3143)(1, 0.3196)(2, 0.3221)(3, 0.3288)(4, 0.3197)};
    \addplot+[color=fig_7_color_2] coordinates{(0, 0.2925)(1, 0.3077)(2, 0.3084)(3, 0.3150)(4, 0.29456)};
    \addplot+[color=fig_7_color_3] coordinates{(0, 0.2600)(1, 0.26288)(2, 0.26328)(3, 0.26920)(4, 0.25811)};
    \addplot+[color=fig_7_color_4] coordinates{(0, 0.1943)(1, 0.1996)(2, 0.2037)(3, 0.2072)(4, 0.1997)};
\end{axis}
\end{tikzpicture}
\\
\ref{named}
\captionsetup{skip=1pt}
\caption{NMI according to different $\lambda$'s in \textsf{META-CODE} on Facebook and Engineering when different percentages of nodes are queried.}
\label{fig:RQ5_lambda}
\end{figure}

\subsubsection{Computational Complexity ({\bf RQ8})} To empirically validate the scalability of our \textsf{META-CODE} method, we measure the execution time on synthetic networks generated using the Erdős–Rényi model \cite{erdHos1960evolution} with a fixed number of nodes and a varying number of edges. The Erdős–Rényi model $G\left( {n,q} \right) $ can generate various networks by creating edges between two nodes with probabilitiy $q$. We generate a set of networks with $n = 2000 $ and $q = \left\{ {0.002,0.003,0.006,0.008,0.01,0.02,0.03,0.04} \right\} $. Additionally, we randomly generate node metadata due to the fact that the quality of the node metadata does not influence the execution time. Fig. \ref{fig:time} illustrates the execution time (in seconds) of one iteration of \textsf{META-CODE}, including the training time of the GNN and MLP models and query node selection, as the number of edges increases. The dashed line indicates a linear scaling in $\left| \mathcal{E} \right| $, derived from Theorem 3. It can be seen that our empirical evaluation concurs with the theoretical analysis.

\begin{figure}[t!]
\pgfplotsset{footnotesize,samples=10}
\centering
\begin{tikzpicture}
    \begin{axis}[        xlabel=$|\mathcal{E}|$,        ylabel= Execution time (s), grid=major, grid style={dashed},  width = 5.95cm, height = 4.25cm , legend style={at={(0.39,0.2)},anchor=west}]
        \addplot [color=fig_8_color_1, mark=square, legend = \textsf{META-CODE}]
            coordinates {
                (1734, 197.2833)(3487, 224.2517)(7021, 276.9718)(10187, 304.2664)(13745, 348.728)(17231, 404.77)(26873, 515.596)(33317, 601.8297)(36206, 629.718)(48505, 796.292)(57380, 890.4584)(62948, 945.412)
            };
        \addplot [ dashed, color=red,        mark=triangle, legend = $O(|\mathcal{E}|)$ ]
            coordinates {
                (1000,188) (11000,310) (21000,433) (31000,554) (41000, 677) (51000,799) (65000,970)
            };
    \addlegendentry{\textsf{META-CODE}}
    \addlegendentry{$O(|\mathcal{E}|)$}
    \end{axis}
\end{tikzpicture}
\captionsetup{skip=2pt}
\caption{The computational complexity of \textsf{META-CODE}, where the plot of the execution time versus $|\mathcal{E}|$ is shown.}
\label{fig:time}
\vspace{-1.0em}
\end{figure}

\begin{figure}[t!]
    \label{fig:visual}
    \begin{center}
        \includegraphics[width=0.9\linewidth]{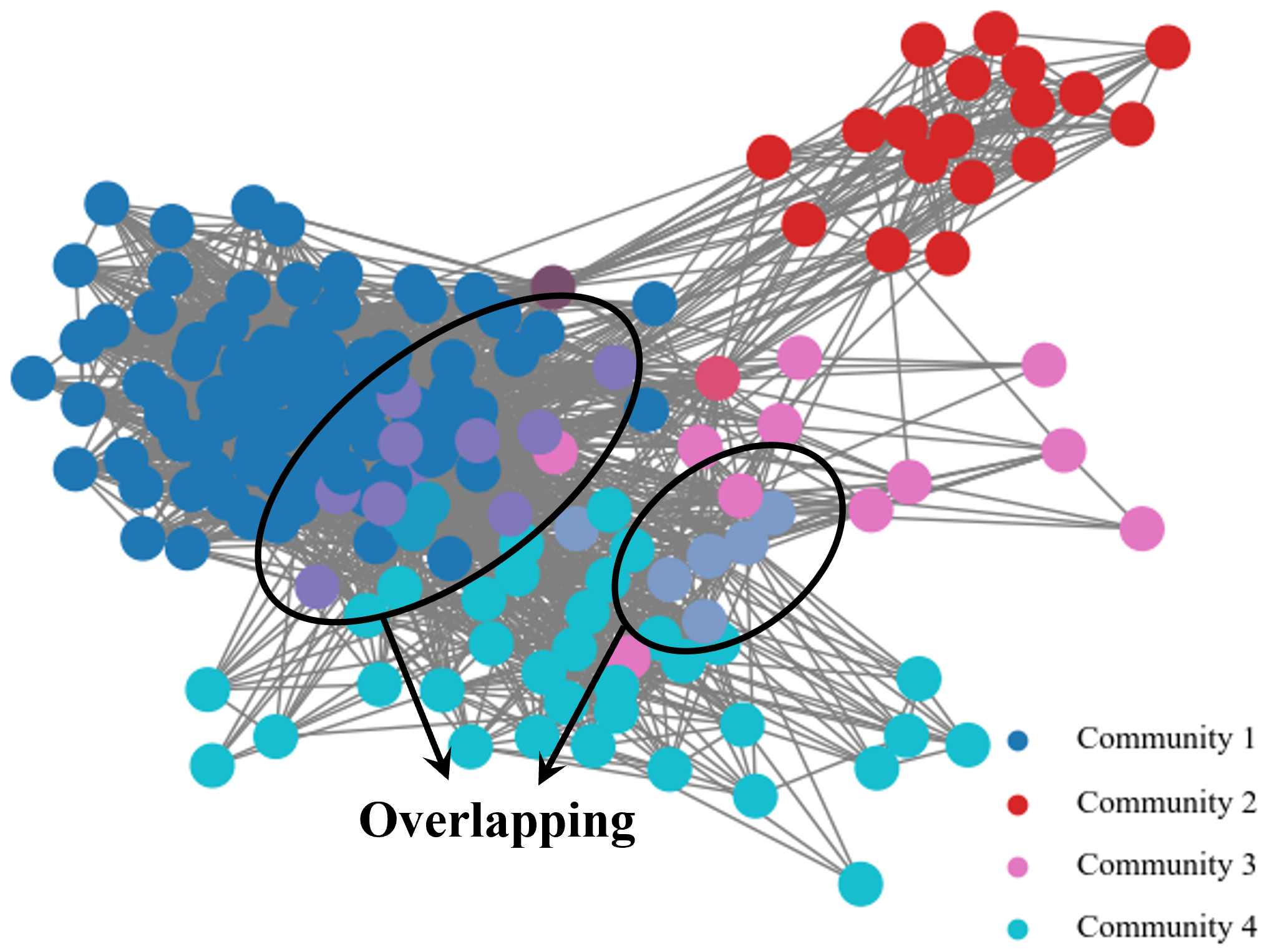}
    \end{center}
    \captionsetup{skip=-10pt}
    \caption{A visualization of overlapping community detection results on the Facebook dataset.}
\vspace{-1.0em}
\end{figure}

\subsubsection{Visualization ({\bf RQ9})}

For clarity and better visualization, we focus on the four largest detected overlapping communities produced by \textsf{META-CODE} on the Facebook dataset and visualize them using a color-coded graph, which is shown in Fig. 9. This selective approach allows us to highlight the key community structures without overwhelming the visualization with smaller or less significant communities. The visualization effectively illustrates how nodes within overlapping communities are distributed and interconnected, showcasing the capability of our method to uncover meaningful and interpretable community structures.

\section{Concluding Remarks}
In this paper, we solved the unexplored problem of overlapping community detection in networks with unknown topology. To tackle this challenge, we introduced \textsf{META-CODE}, a novel unified framework for overlapping community detection, which iteratively performs the following three steps: 1) community-affiliation embedding, 2) network exploration via node queries, 3) network inference. Through extensive experiments on three real-world datasets, we demonstrated (a) the superiority of \textsf{META-CODE} over state-of-the-art overlapping community detection methods, achieving dramatic gains up to 65.55\% in terms of NMI compared to the best existing competitor, (b) the impact of each module in \textsf{META-CODE}, and (c) the effectiveness of our node query strategy by analyzing the growth of the explored subnetworks. Our theoretical analyses also uncovered (a) the effect of the proposed node query strategy and (b) the scalability of \textsf{META-CODE}.

\section*{Appendix}

\subsection{Proof of Theorem 1}

Let us denote the $i$-th community to which a node belongs as $C_i$. Since the AGM is used to generate a network $\mathcal{G}$ by creating edges between nodes in each community independently, the degree distribution of a node in a given community $C_i$, denoted by $Y_{C_i}$, follows the binomial distribution  
\begin{equation}
Y_{C_i }  \sim B\left( {N_{C_i }  - 1,p} \right),
\end{equation}
where $N_{C_i}$ is the number of nodes in $C_i$. By the properties of independence and additivity, the degree distribution $\mathcal{D}_M$ (resp. $\mathcal{D}_M'$) of any node $u$ (resp. $v$) belonging to $M$ (resp. $M'$) communities still follows the binomial distribution. Thus, we have
\begin{equation}
\begin{array}{l}
 \mathbb{E}_u \left[ {\mathcal{D}_M } \right] - \mathbb{E}_v \left[ {\mathcal{D}_{M'} } \right] \\ = \sum\limits_{i = 1}^M {\left( {N_{C_i }  - 1} \right)} p - \sum\limits_{i = 1}^{M'} {\left( {N_{C_i }  - 1} \right)} p \\ 
  \ge M\left( {N_{\min }  - 1} \right)p - M'\left( {N_{\max }  - 1} \right)p \\ 
 \ge - M'\varepsilon p + \left( {M - M'} \right)\left( {N_{\min }  - 1} \right)p \\ 
 \ge - M'\left(\frac{{N_{\min }  - 1}}{K}-1\right)p + \left( {M - M'} \right)\left( {N_{\min }  - 1} \right)p \\ 
  = \left( {M - M' - \frac{{M'}}{K}} \right)\left( {N_{\min }  - 1} \right)p + M'p \\ 
  \ge 0, \\ 
 \end{array}
\end{equation}
where the second inequality holds due to Assumption 2 and the last inequality comes from $M'<M\le K$. This completes the proof of this theorem.

\subsection{Proof of Theorem 2}

The average degree of a randomly selected node $v$ in $\mathcal{G}$, denoted by $\mathbb{E}_v \left[ \mathcal{D} \right]$, is given by
\begin{equation}
\begin{array}{l}
 \mathbb{E}_v \left[ \mathcal{D} \right] = \sum\limits_{i = 1}^K {\mathbb{E}_v } \left[ {\mathcal{D}\left| {A_i } \right.} \right]\mathbb{P}\left( {A_i } \right) \\ 
 \;\;\;\;\;\;\;\;\;\;\; = \frac{{\sum\limits_{i = 1}^K {\mathbb{E}_v \left[ {\mathcal{D}\left| {A_i } \right.} \right]N_i } }}{N} \\
 \;\;\;\;\;\;\;\;\;\;\;\le  \frac{2\mathbb{E}_v \left[ {\mathcal{D}\left| {A_1 } \right.} \right]N_1 }{N} \\ 
\;\;\;\;\;\;\;\;\;\;\;\le 2\left(N_{\max }-1\right) p\frac{K}{{K + 1}} \\
\;\;\;\;\;\;\;\;\;\;\;\le 2\left( {N_{\max }  - 1} \right)p\frac{{N_{\min }  - 1}}{{N_{\min }  + \varepsilon }} \\
\;\;\;\;\;\;\;\;\;\;\; \le 2\left(N_{\max }-1\right) p\frac{{N_{\min } - 1}}{{N_{\max } }} \\ 
\;\;\;\;\;\;\;\;\;\;\; \le \sum\limits_{i = 1}^M {\left( {N_{C_i }  - 1} \right)} p \\ 
\;\;\;\;\;\;\;\;\;\;\; =\mathbb{E}_u\left[\mathcal{D}_M \right],
 \end{array}
\end{equation}
where the first two inequalities hold due to Assumption 3 and the third and fourth inequalities come from $\varepsilon\le\frac{N_\text{min}-1}{K}-1$ and Assumption 2, respectively. This completes the proof of this theorem.

\subsection{Proof of Theorem 3}

We analyze the computational complexity of each step in \textsf{META-CODE}. In Step 1, a GNN model is utilized to generate the community-affiliation embedding matrix ${\bf F}$, with a training complexity of $\mathcal{O}\left( {L_1\left| \mathcal{E} \right|} \right) $ per epoch \cite{wu2020comprehensive}, where $L_1$ is the number of GNN layers. In Step 2, the node query strategy in (9) of the main manuscript has a complexity of $\mathcal{O}\left( {NK + tNK} \right) $, where $N$ and $K$ are the number of nodes and communities, respectively. Finally, in Step 3, the training of MLPs dominates the complexity of this step and has a complexity of $\mathcal{O}\left( {L_2\left| \mathcal{E} \right|} \right) $ \cite{scikit-learn},  where $L_2$ is the number of MLP layers, conditioned that the size of model parameters ${\left| {{\bf W}_{Sia} } \right|} $ in MLPs does not scale with ${\left| \mathcal{E} \right|} $. Therefore, the total complexity of our \textsf{META-CODE} method is finally bounded by $\mathcal{O}\left( {\left| \mathcal{E} \right|} \right) $. This completes the proof of this theorem.

\section*{Acknowledgments}
The work of W.-Y. Shin was supported by the National Research Foundation of Korea (NRF) grant funded by Korea government (MSIT) under Grants RS-2021-NR059723 and RS-2023-00220762. Ming Li acknowledged the support from the Jinhua Science and Technology Plan (No. 2023-3-003a).

\bibliographystyle{IEEEtran}
\bibliography{bibfile}

\end{document}